\documentclass[reprint,aps,prb,onecolumn,groupedaddress,nobibnotes]{revtex4-2}
\usepackage{amssymb}
\usepackage{graphicx}
\usepackage[caption=false]{subfig} % for figure caption look better :)
\usepackage{float}
\usepackage{amsmath}
\usepackage{dcolumn}% Align table columns on decimal point
\usepackage{color}
\usepackage{graphicx}
\usepackage[pdftex,colorlinks=true,linkcolor=red,citecolor=blue]{hyperref}
\usepackage{soul}

\usepackage{setspace}
\begin{document}

\title[]{Fermiology of the kagome compound LuNb$_6$Sn$_6$ probed by de Haas–van Alphen oscillations}
\author{Tucker Beekmann$^{1}$}
\altaffiliation{Equal contribution}
\author{Caue Kaufmann Ribeiro$^{2}$}
\altaffiliation{Equal contribution}
\author{Kyryl Shtefiienko$^{1}$}
\author{Jiaqiang Yan$^{3}$}
\author{Brenden R. Ortiz$^{3}$}
\author{Christopher A. Mizzi$^{2}$}
\author{Keshav Shrestha$^{1}$}
\email{E-mail: kshrestha@wtamu.edu}

\affiliation{$^{1}$Department of Chemistry and Physics, West Texas A$\&$M University, Canyon, Texas 79016, USA}
\affiliation{$^{2}$National High Magnetic Field Laboratory, Los Alamos National Laboratory, Los Alamos, New Mexico 87545, USA}
\affiliation{$^{3}$Materials Science and Technology Division, Oak Ridge National Laboratory, Oak Ridge,Tennessee 37831, USA}

\begin{abstract}
We report a detailed de Haas--van Alphen (dHvA) study of the recently discovered kagome metal LuNb$_6$Sn$_6$ using torque magnetometry, magnetization, and heat-capacity measurements. Temperature-dependent torque and heat-capacity data reveal a charge density wave (CDW) transition at $T_{\mathrm{CDW}} = 85$ K. The thermal hysteresis observed in both measurements establishes the first-order nature of the transition. Quantum oscillation measurements identify two major dHvA frequencies: $F_\alpha \approx 20$~T and $F_\beta \approx 200$~T, and their angular dependence is consistent with ellipsoidal Fermi surface (FS) pockets. Landau fan diagram analysis reveals evidence for a nontrivial Berry phase associated with the $F_\alpha$ pocket, indicating possible nontrivial electronic topology in LuNb$_6$Sn$_6$. Analysis of the temperature and magnetic field dependence of the oscillations using the Lifshitz–Kosevich formula yields electronic parameters that indicate anisotropic quantum transport properties. First-principles calculations provide further insight into the electronic structure, revealing Dirac-like band crossings, a flat band, and multiple van Hove singularities near the Fermi level. Our calculations based on the pristine phase cannot fully reproduce the experimentally observed quantum oscillation frequencies, suggesting that CDW-induced FS reconstruction plays a crucial role in the ground-state electronic structure of LuNb$_6$Sn$_6$. These results provide new insight into the FS topology and electronic structure of LuNb$_6$Sn$_6$, enriching our understanding of the electronic properties of kagome materials.
\end{abstract}

\pacs{}

\maketitle

\section{Introduction}
Kagome materials have recently attracted significant attention in condensed matter physics due to their unique geometric lattice and emergent electronic and topological properties~\cite{yin2022topological,wilson2024v3sb5,wang2023quantum,shtefiienko2025electronic}. These compounds crystallize in a two-dimensional network of corner-sharing triangles and hexagons, resembling the traditional Japanese bamboo basket pattern~\cite{mekata2003kagome}. The kagome lattice naturally hosts flat band (FB), Dirac points (DPs), and van Hove singularities (VHSs), often located near the Fermi level ($E_F$), making these systems highly susceptible to electronic instabilities and correlated quantum phenomena~\cite{peng2021realizing,bhandari2025pressure,shrestha2023electronic}. As a result, kagome materials exhibit a variety of exotic states, including charge density wave (CDW) order and superconductivity~\cite{arachchige2022charge,werhahn2022kagome,ortiz2023evolution,ortiz2024intricate,phillips2025electrical,ye2018massive,liu2018giant,li2025giant}.

As these electronic features lie close to the $E_F$, external tuning parameters such as chemical substitution~\cite{hou2023effect,yang2022titanium,oey2022fermi,kautzsch2023incommensurate}, strain~\cite{qian2021revealing,lin2024uniaxial,yang2023plane}, or applied pressure~\cite{yu2021unusual,chen2021double,phillips2024fermi} can effectively drive novel quantum phases. Several recent review articles have highlighted the structural diversity, electronic correlations, and topological characteristics of kagome compounds, emphasizing their role as a fertile platform for discovering emergent quantum states~\cite{di2026kagome,hu2023electronic,wang2023quantum,wang2024topological,jiang2023kagome,yin2022topological,bernevig2022progress}. One of the families, $A$V$_3$Sb$_5$ ($A$ = K, Rb, Cs) exhibit both superconductivity ($T_c$ $\sim$ 0.9 to 3 K) and CDW order at $T_{\mathrm{CDW}} \sim 80$--100 K~\cite{ortiz2019new,ortiz2020cs}. Recent quantum oscillation studies~\cite{yu2021concurrence, yin2021superconductivity,shrestha2023high,yang2020giant,nakayama2022carrier,ortiz2021fermi,shrestha2023fermi,luo2022electronic,fu2021quantum,zhang2022emergence,broyles2022effect,shrestha2022nontrivial,shrestha2023electronic,rosenberg2024probing,phillips2025electrical,shtefiienko2025electronic,zheng2024quantum,yi2024quantum} have further elucidated the Fermi surface (FS) topology of $A$V$_3$Sb$_5$, confirming signatures of nontrivial band topology and revealing substantial FS reconstruction associated with the CDW phase.

LuNb$_6$Sn$_6$ is a member of the recently discovered $Ln$Nb$_6$Sn$_6$ ($Ln$ = Ce–Lu, Y) family of kagome metals~\cite{ortiz2025stability}. Structurally, it is analogous to the well-studied $R$V$_6$Sn$_6$ ($R$ = Sc, Y, or rare earth) systems, also known as the 1-6-6 family, and crystallizes in the HfFe$_6$Ge$_6$-type structure (space group \textit{P6/mmm}), a member of the broader $AM_6X_6$ materials class~\cite{ortiz2025stability,lou2025orbital,ingham2024theory}. The structure consists of corner-sharing kagome layers of Nb atoms stacked along the $c$ axis, as shown in Fig.~\ref{Fig1}(a). Bulk measurements reveal a CDW transition near 68 K, and single-crystal x-ray scattering identifies a $\sqrt{3} \times \sqrt{3} \times 3$ structural modulation in the CDW phase~\cite{ortiz2025stability}. Recently, scanning tunneling microscopy/spectroscopy and first-principles calculations proposed a FS frustration scenario, in which competing FS instabilities suppress the dominant CDW ordering tendency and lead to the emergence of a secondary CDW state~\cite{yang2025fermi}. Under applied pressure, the CDW at 68 K is gradually suppressed and disappears near 2 GPa, although no superconducting phase has been reported~\cite{meier2025pressure}.

Despite these intriguing collective phenomena, the electronic structure of LuNb$_6$Sn$_6$ remains largely unexplored. In particular, quantum oscillation measurements, which provide direct access to the FS geometry and quasiparticle effective masses, have not yet been reported for this compound. Recently, improvements in the crystal growth method~\cite{yan2017flux} have produced LuNb$_6$Sn$_6$ single crystals exhibiting a substantially enhanced $T_{\mathrm{CDW}}$. In this work, we investigate both the Fermiology and the enhanced CDW properties of LuNb$_6$Sn$_6$ by combining comprehensive de Haas--van Alphen (dHvA) measurements using torque magnetometry and vibrating-sample magnetometry (VSM) with heat-capacity measurements. The dHvA measurements reveal the FS properties of the material, while heat capacity and torque provides a probe to identify the CDW transition. Our torque and magnetization data in magnetic fields up to 14 T reveal clear oscillations with two dominant frequencies: $F_\alpha \approx 20$ T and $F_\beta \approx 200$ T, when the field is applied along the $ab$ plane. Detailed angle-dependent dHvA measurements further show that these frequencies originate from ellipsoidal Fermi pockets of LuNb$_6$Sn$_6$. We analyze the temperature- and field-dependent oscillation amplitudes using the Lifshitz–Kosevich (LK) formula and extract key physical parameters characterizing the electronic properties of this material. 

\section{Experimental and Computational details}
High-quality single crystals of LuNb$_6$Sn$_6$ were grown using a Sn-flux method in a horizontal configuration~\cite{yan2017flux}. The starting elements were mixed in a nominal molar ratio of 2:1:20 and sealed in a 6~in.-long, 0.5~in.-diameter quartz tube. High-purity Lu metal (Ames), Nb shot (Alfa, 99.99\%), and Sn shot (99.999\%) were used as starting materials. The ampoule was sealed under an argon atmosphere and placed in a single-zone horizontal furnace. The Nb slugs were positioned at the hot end of the furnace, which was maintained at $950\,^{\circ}\mathrm{C}$. A natural temperature gradient of approximately $40\,^{\circ}\mathrm{C}$ was established along the length of the ampoule. After three weeks, the furnace was powered off and allowed to cool naturally to room temperature. The tube was then opened, the flux was resealed in quartz, and centrifuged to remove excess Sn flux. The resulting crystals can reach masses of up to 250~mg and typical dimensions of approximately $1 \times 1 \times 0.1$~cm$^3$ in hexagonal plate form, representing more than two orders of magnitude improvement over our original synthesis method~\cite{ortiz2025stability}.

The phase purity and composition of the crystals were examined by powder X-ray diffraction (XRD), energy-dispersive spectroscopy (EDS), and SEM-EDS elemental mapping. The powder XRD pattern is consistent with the crystal structure of LuNb$_6$Sn$_6$ and shows no evidence of secondary phases. The elemental maps reveal a homogeneous distribution of Lu, Nb, and Sn throughout the crystal, and the measured composition is consistent with the expected 1:6:6 stoichiometry.

Torque and magnetization measurements were carried using the standard torque and VSM options of a 14 T Physical Property Measurement System (PPMS Dynacool, Quantum Design). For torque measurements, the sample platform was mounted on a horizontal rotator to enable angle-dependent measurements. For the VSM measurements, the sample was mounted on a quartz rod with GE varnish with the magnetic field applied parallel to the $ab$ plane and the $c$ axis. Note: owing to the small magnetization signal from the sample, there may be non-negligible background contributions from the quartz and GE varnish to the measured magnetic moment. While these will not affect the quantum oscillation amplitudes, they may lead to inaccurate absolute magnetic moment values.

Heat capacity was measured using the commercial micro-calorimeter from Quantum Design. In addition to the standard 2$\tau$ relaxation method with a 2\% temperature rise, a large pulse single-slope method was employed with a 20--30\% rise to examine differences between heating and cooling curves in the vicinity of the CDW phase transition.

Density functional theory (DFT) calculations were performed using Quantum ESPRESSO (QE)~\cite{QE-2017,QE-2009} with the Perdew--Burke--Ernzerhof (PBE) generalized gradient approximation~\cite{perdew1996generalized,perdew2008restoring} and PAW pseudopotentials from the PSLibrary. All calculations were performed using the pristine phase of LuNb$_6$Sn$_6$. The crystal structure was optimized using the \texttt{vc-relax} option in QE prior to the electronic structure calculations. Plane-wave and charge-density cutoffs of 70 and 560 Ry, respectively, were employed. The self-consistent calculations used a $20\times20\times15$ Monkhorst--Pack $k$-point mesh~\cite{monkhorst1976special} with Gaussian smearing and a convergence threshold of $10^{-9}$ Ry. FS calculations were carried out using a denser $40\times40\times40$ $k$-point mesh.

\section{Results and discussion}
Figure~\ref{Fig1}(b) displays the temperature dependence of the torque ($\tau$) for a LuNb$_6$Sn$_6$ single crystal (S1) measured under an applied field of 14 T from 2 K to room temperature at a fixed angle. The torque signal shows a clear anomaly at $T_{\mathrm{CDW}} = 85$ K, as indicated by the arrow, corresponding to the CDW transition in LuNb$_6$Sn$_6$. As shown in the upper inset, the CDW transition exhibits a hysteresis of about 1 K between the cooling and warming curves. The hysteresis observed in the torque data is consistent with the behavior reported in previous heat-capacity measurements~\cite{ortiz2025stability}.

It is important to note that the observed transition temperature is significantly higher than the values reported previously from resistivity, magnetization, and heat-capacity measurements ($T_{\mathrm{CDW}} \approx 68$ K)~\cite{ortiz2025stability} and temperature-dependent X-ray diffuse scattering ($T_{\mathrm{CDW}} = 70$ K)~\cite{yang2025fermi}. To verify the higher transition temperature observed in the torque measurements and confirm the first order nature of the transition, we performed heat-capacity measurements on the same sample (S1), as shown in Fig.~\ref{Fig1}(c). 
The heat-capacity data exhibit a pronounced anomaly at $T_{\mathrm{CDW}} = 84.3$ K, in excellent agreement with the transition temperature determined from the torque data [Fig.~\ref{Fig1}(b)]. Furthermore, the thermal hysteresis observed between the cooling and heating heat-capacity curves [Fig.~\ref{Fig1}(c), inset] is consistent with both the torque measurements [Fig.~\ref{Fig1}(b), inset] and previous heat-capacity studies~\cite{ortiz2025stability}.

The higher CDW transition temperature observed in our sample may be related to reduced disorder resulting from the improved crystal-growth process. However, a detailed understanding of the relationship between sample quality and the CDW transition in LuNb$_6$Sn$_6$ requires further investigation. Furthermore, a very weak hump-like feature is observed in the torque data near $\sim$ 74 K; however, no corresponding anomaly is detected in the heat-capacity data [Fig.~\ref{Fig1}(c)], suggesting that this feature is unlikely to represent an intrinsic bulk phase transition. Since this tempearature range lies close to the liquid-nitrogen condensation temperature ($\sim 77~\mathrm{K}$ at atmospheric pressure), the feature is likely an experimental artifact.

To probe the FS of LuNb$_6$Sn$_6$, we measured the torque as a function of applied magnetic field up to 14 T. Figure~\ref{Fig2}(a) shows the torque versus magnetic field at 2 K with a tilt angle $\theta = 45^\circ$. Here, $\theta$ is defined as the angle between the $c$ axis and the magnetic field direction for rotations from the $c$ axis to the $ab$ plane, as illustrated in the inset. As shown in Fig.~\ref{Fig2}(a), clear dHvA oscillations appear in the torque signal above 2 T. To isolate the oscillatory component, we subtracted a smooth polynomial background from the data; because torque goes as $H^2$ to lowest order, the solid red curve represents a second-order polynomial fit. The background-subtracted signal, $\Delta\tau$, is shown in Fig.~\ref{Fig2}(b). Multiple oscillation periods are visible, indicating more than one frequency, which is further confirmed by the fast Fourier transform (FFT) of the data.

Figure~\ref{Fig2}(c) shows the frequency spectrum of the dHvA oscillations obtained from the FFT of the oscillatory signal. Two major peaks are observed at $F_\alpha = 12.8$ T and $F_\beta = 259$ T. The peak near 25 T corresponds to the second harmonic of $F_\alpha$. A small peak around 41 T is also observed at a few angles; however, the amplitude of this peak is weak, and the signal is largely dominated by $F_\alpha$ and $F_\beta$. Therefore, this frequency is not considered in the present analysis. According to the Onsager relation~\cite{Shoenberg, ando2013topological,shrestha2014shubnikov,shrestha2018evidence}, the quantum oscillation frequency $F$ is directly proportional to the extremal cross-sectional area $A_F$ of the FS perpendicular to the applied magnetic field, expressed as $F = (\hbar/2\pi e)A_F$, where $\hbar$ is the reduced Planck constant and $e$ is the elementary charge. Therefore, measuring dHvA oscillations at different tilt angles and tracking the angular evolution of $F_\alpha$ and $F_\beta$ provide important information about the dimensionality and topology of the corresponding FS pockets. To investigate this in LuNb$_6$Sn$_6$, torque measurements were carried out at different tilt angles $\theta$, with the magnetic field applied along various crystallographic directions. 

Figure~\ref{Fig3}(a) presents the torque versus magnetic field data for several $\theta$ values. A clear change in the oscillation pattern is observed as $\theta$ varies. To extract the dHvA oscillatory component at each angle, a smooth second-order polynomial background was subtracted from the raw torque data, followed by a FFT analysis to determine the oscillation frequencies, similar to the procedure used in Fig.~\ref{Fig2}. The resulting frequency spectra of the dHvA oscillations at different $\theta$ values are shown in Figs.~\ref{Fig3}(b) and (c) for $F_\alpha$ and $F_\beta$, respectively. As indicated by the dashed arrows in both panels, both frequencies exhibit clear angle dependence. In addition, a few weak peaks appear at certain angles. However, these frequencies could not be reliably tracked at higher angles; therefore, they are not discussed further in this manuscript. We note that torque measurements on other $R$V$_6$Sn$_6$ ($R$ = rare earth) kagome compounds~\cite{zheng2024quantum,yi2024quantum,shtefiienko2025electronic,phillips2025electrical} similarly report two major oscillation frequencies that remain resolvable as the sample is rotated through different $\theta$ values.

For a quantitative analysis, $F_\alpha$ and $F_\beta$ were extracted at each tilt angle and plotted as a function of $\theta$, as shown in Fig.~\ref{Fig4}. For comparison, the frequency values obtained from VSM measurements at $H \parallel c$ ($\theta = 0^\circ$) and $H \parallel ab$ ($\theta = 90^\circ$) are also included in both panels, and are consistent with those derived from the torque measurements. As evident from Fig.~\ref{Fig4}, the angle dependence of $F_\alpha$ and $F_\beta$ deviates from that expected for either a cylindrical or a spherical FS. Previous quantum oscillation measurements and first-principles calculations on related kagome compounds~\cite{zhang2022emergence,rehfuss2024quantum,shrestha2022nontrivial,shrestha2023electronic} have revealed quasi-two-dimensional FS sheets. In contrast, the present angle dependences are consistent with ellipsoidal FS pockets (dashed curves).

To quantify the observed angle dependence, we model the frequency--angle relation using an ellipsoidal FS~\cite{campbell2017quantum,sebastian2008quantum}, $F(\theta) = \frac{F_0}{\sqrt{\cos^2 \theta + \frac{1}{\epsilon}\sin^2 \theta}}$,
where $F_0$ is the frequency at $\theta = 0^\circ$ and $\epsilon$ characterizes the anisotropy of the FS. As shown by the dashed curves in Fig.~\ref{Fig4}, this model provides an excellent description of the angle dependence of both $F_\alpha$ and $F_\beta$. The best fits yield $F_0 = (10.14 \pm 0.11)$~T and $\epsilon = (4.08 \pm 0.15)$ for $F_\alpha$, and $F_0 = (388.82 \pm 2.24)$~T and $\epsilon = (0.27 \pm 0.01)$ for $F_\beta$. These results indicate that both frequencies originate from anisotropic, three-dimensional Fermi pockets of LuNb$_6$Sn$_6$.

To confirm the dHvA oscillations observed in the torque measurements of crystal S1, magnetization measurements were performed on three single crystals (S1, S2, and S3). All three samples exhibit clear dHvA oscillations above 2 T. The oscillations are even clearer in the background-subtracted data (background approximated with a second order polynomial), as presented in Fig.~\ref{Fig5}(a). The presence of beating patterns indicates multiple oscillation frequencies, as confirmed by the FFT analysis presented in Fig.~\ref{Fig5}(b). Two dominant frequencies, $F_\alpha = 19$~T and $F_\beta = 204$~T, are observed in all samples and are in good agreement with those obtained from the torque data at $\theta = 90^\circ$ (Fig.~\ref{Fig4}). These results demonstrate good sample homogeneity and confirm the reproducibility and intrinsic nature of the observed quantum oscillations in LuNb$_6$Sn$_6$, including the presence of the two dominant frequencies, $F_\alpha$ and $F_\beta$.

As demonstrated above, our detailed torque and magnetization measurements of LuNb$_6$Sn$_6$ under magnetic fields reveal the presence of two major frequencies, $F_\alpha$ and $F_\beta$, in the dHvA oscillations and confirm that they originate from ellipsoidal Fermi pockets. To extract the relevant physical parameters of these Fermi pockets, we performed dHvA measurements at different temperatures. Figures~\ref{Fig6}(a) and (b) display the background-subtracted magnetization as a function of magnetic field for $H \parallel ab$ and $H \parallel c$, respectively, measured at various temperatures. Clear dHvA oscillations are observed for both field orientations, and the corresponding frequency spectra are shown in Figs.~\ref{Fig6}(c) and (d). The oscillations disappear at higher temperatures and are barely visible above 10 K for both field directions.

The amplitude of the dHvA oscillations gradually decreases with increasing temperature and decreasing field, as also reflected in the corresponding FFT spectra [Figs.~\ref{Fig6}(c) and (d)]. The temperature- and field-dependent attenuation of the oscillation amplitude is well described by the LK formalism~\cite{Shoenberg}. According to the LK formula, the dHvA oscillations is given by

\begin{equation}
\Delta M (T, H) \propto e^{-\lambda_D (H)} {\lambda(T/H)\over {sinh[\lambda(T/H)]}},\label{LK}
\end{equation}
with $\lambda_D (H) = {2\pi^2k_B\over {\hbar e}} m^{*} {T_D\over H}$ and
$\lambda(T/H)={2\pi^2k_B\over {\hbar e}} m^{*} {T\over H}$. Here, $T_D$, $k_B$ and $m^*$ represent the Dingle temperature, Boltzmann's constant, and effective mass of the charge carriers, respectively. The first term is the Dingle factor, which describes the damping of the quantum oscillation amplitude due to impurity scattering. The second term accounts for the thermal damping caused by the thermal broadening of the Fermi-Dirac distribution.

The temperature dependence of the FFT amplitudes for the $\alpha$ and $\beta$ orbits for $H \parallel ab$ and $H \parallel c$ are shown in Figs.~\ref{Fig7}(a) and (b), respectively. For $H \parallel c$, the $F_\beta$ peak is relatively weak [inset of Fig.~\ref{Fig7}(b)] and can only be observed above 10 T. The solid curves in Figs.~\ref{Fig7}(a) and (b) represent the Lifshitz--Kosevich (LK) fits using Eq.~\ref{LK}, where the magnetic field was taken as the average field over the FFT window.

Focusing on sample S3 because the magnetization was measured in both directions in this sample, we extract effective cyclotron masses from the LK fitting of $m^{*}_\alpha$ = 0.088(4)$m_o$ and $m^{*}_\beta$ = 0.28(1)$m_o$ for $H \parallel ab$, and $m^{*}_{\alpha}$ = 0.052(1)$m_o$ and $m^{*}_\beta$ = 0.22(2)$m_o$ for $H \parallel c$, where $m_o$ is the rest mass of an electron. Uncertainties in the masses were directly obtained from the fitting procedure. The small values of $m^{*}$ indicate light charge carriers and are comparable to those reported for other kagome metals~\cite{shrestha2022nontrivial,dhital2024fermi,phillips2024fermi}.

To extract additional physical parameters, we evaluated $T_D$ from the magnetic-field dependence of the dHvA oscillation amplitude at 2 K. The full form of the LK expression was used to perform this fit with the effective mass fixed to the value determined from the temperature-dependent LK amplitude shown in Fig.~\ref{Fig7}. The insets of both subfigures display the results of these fittings. From the fitting parameters, we obtained $T_D = 8.22(4)$ K and $2.95(7)$ K for the $\alpha$ and $\beta$ orbits for $H \parallel ab$, respectively, and $T_D =4.03(1)$ K for the $\alpha$ orbit for $H \parallel c$.  Uncertainties in $T_D$ were also found from the fitting procedure. The oscillation amplitude of the $\beta$ orbit was too weak to reliably extract $T_D$.

Using the values of  $F$, $m^{*}$, and $T_D$, we further determined several relevant physical parameters for the charge carriers. The uncertainties in all  parameters listed in Table I were obtained by standard error propagation assuming $F$, $m^*$, and $T_D$ are independent variables. Taking the $F_\alpha = 19.75(7)$ T orbit of sample S3 as a representative example, the Fermi wave vector was obtained from the Onsager relation, $F = (\hbar/2e)k_F^2$, yielding $k_F = 2.450(4) \times 10^{-2}$ \AA$^{-1}$. The Fermi velocity was estimated using $v_F = \hbar k_F / m^{*}$, giving $v_F = 3.22(2) \times 10^{5}$ m.s$^{-1}$. Using $T_D = 8.22(4)$ K, the quantum lifetime was calculated from $\tau_q = \hbar/(2\pi k_B T_D)$, resulting in $\tau_q = 0.148(6)$ ps. The corresponding mean free path, $\ell = v_F \tau_q$, and carrier mobility, $\mu = e \tau_q / m^{*}$, were estimated to be $48(3)$ nm and $2960(200)$ cm$^{2}.$V$^{-1}$.s$^{-1}$, respectively. These values indicate relatively high carrier mobility and coherent quasiparticle transport. The same procedure was applied to determine the physical parameters for the remaining samples, orientations, and frequencies.

The $m^*$ and $T_D$ for the $\alpha$ orbit exhibit clear anisotropy, with larger values for $H \parallel ab$ than for $H \parallel c$. The anisotropy in the effective mass likely reflects the anisotropic FS as evidenced by the anisotropy in the quantum oscillation frequencies in the two directions. The $T_D$ anisotropy is slightly larger than the mass anisotropy which is attributed to the difference  in $\ell$  for $H \parallel ab$ and $H \parallel c$, resulting in a quantum mobility nearly $3.5\times$ larger for $H \parallel c$.  Such anisotropic transport is characteristic of layered materials and has been reported in kagome systems such as Fe$_3$Sn~\cite{prodan2024anisotropic} and FeGe~\cite{ma2025anisotropic}, as well as in ZrSiS~\cite{hu2017nearly,miertschin2024anisotropic}. In contrast, the $\beta$ orbit exhibits a quantum oscillation frequency for $H || c$  that is nearly twice as large as the frequency for $H || ab$, while the effective mass only changes by $\approx30\%$ with orientation. 

To obtain information about the topology of the electronic bands observed with quantum oscillations, we analyze the phase associated with the quantum oscillations via a Landau fan diagram. We focus on our magnetization measurements (H $\parallel$ ab) to demonstrate the robustness of the results across different samples. In short, quantum oscillations arise from Landau quantization of the electronic states and manifest in an oscillatory contribution to the thermodynamic potential, $\Omega_{\mathrm{osc}}\propto \cos[2\pi(\frac{F}{B}+\phi)]$, where $F$ is the quantum-oscillation frequency and $\phi$ is the thermodynamic phase. Within the Lifshitz--Onsager quantization condition, $\phi=-1/2+\Phi_B/2\pi+\delta$, where $\Phi_B$ is the Berry phase, the $-1/2$ term originates from the semiclassical Maslov phase, and $\delta$ is the dimensional phase correction. For a three-dimensional FS, such as the ellipsoidal pockets depicted in Fig.~\ref{Fig4}, $\delta=\pm 1/8$~\cite{Shoenberg}. The oscillatory component of the magnetization arising from Landau quantization is given by $M_{\mathrm{osc}}=-\partial\Omega_{\mathrm{osc}}/\partial B$, which yields $M_{\mathrm{osc}}\propto -\sin[2\pi(\frac{F}{B}+\phi)]$. Therefore, magnetization maxima occur when $\frac{F}{B}+\phi=n+1/4$, while minima occur when $\frac{F}{B}+\phi=n-1/4$, where n is the LL index. As a result, the oscillation extrema can be fitted using a Landau fan diagram to extract the phase $\phi$ from the intercept. Then, based upon an assumed value of $\delta$, the Berry phase can be determined.

By plotting $n$, as a function of $1/\mu_0H$ and fitting the data using $n = \frac{F}{B}+\phi$, we extract the phase $\phi$, as shown in Fig. 8. For the low-frequency pocket, the extracted phases are inconsistent with the values expected for a trivial Berry phase, $\Phi_B=0$ or $2\pi$, indicated by the horizontal gray lines corresponding to $\delta=\pm1/8$. Instead, all three samples consistently yield a phase shifted from the trivial values, demonstrating the presence of a nontrivial Berry phase. Averaging the results from the three samples gives $\phi=-0.2(1)$, which corresponds to a Berry phase of $\Phi_B=0.85(2)\pi$ or $\Phi_B=0.35(2)\pi$  for $\delta = -1/8$ and $\delta = +1/8$, respectively. In contrast, the high-frequency pocket yields an average phase consistent with the trivial value $\phi=1/8$, corresponding to $\delta=+1/8$ and $\Phi_B=0$. Although assuming $\delta=-1/8$ yields a Berry phase of $\Phi_B=0.5\pi$, the topological character of this pocket remains inconclusive. We therefore leave its definitive determination to future work, where complementary measurements, such as Hall-effect experiments, will be necessary to establish the nature of the charge carriers and the associated band topology. The excellent agreement among the independently extracted phases for each pocket demonstrates the reproducibility and robustness of the analysis across distinct samples. The frequencies extracted from the slopes of the Landau fan diagrams are $F=19.0(4)$ T and $202(1)$ T in good agreement with the corresponding values obtained from the FFT analysis.

To shed further light on the electronic properties of LuNb$_6$Sn$_6$, we computed the electronic band structure and density of states (DOS) of its pristine phase using the DFT. The electronic band structure along the high-symmetry path (inset) is presented in Fig.~\ref{Fig9}(a). The calculated bands exhibit several notable features, including multiple VHSs, a FB, and a DP, as indicated by the shaded regions, arrows, and dotted circle, respectively. These features are located in close proximity to the $E_F$. For example, one of the VHSs lies at the $E_F$, while the DP is located slightly below it. The FB appears approximately 0.65 eV above the $E_F$. With the inclusion of spin-orbit coupling (SOC), there is little or no effect on the FB and VHSs; however, Dirac-like crossings along the M-K direction show a gap, as shown in Fig.~\ref{Fig9}(c). The DP at the K point shows a gap of $\sim$120 meV. The calculated electronic band structure is consistent with previous reports on LuNb$_6$Sn$_6$ and related $R$V$_6$Sn$_6$ compounds, such as ScV$_6$Sn$_6$, LuV$_6$Sn$_6$, and YV$_6$Sn$_6$~\cite{lou2025orbital,shrestha2023electronic,rosenberg2024probing,phillips2025electrical}.

The total and partial DOS (Fig.~\ref{Fig9}(b)), together with the orbital-resolved electronic band structure, further reveal the electronic characteristics of pristine LuNb$_6$Sn$_6$. The electronic states near the $E_F$ are predominantly derived from Nb-$4d$ orbitals. Although a local minimum in the DOS is observed at the $E_F$, the DOS remains finite, and multiple bands cross the $E_F$, consistent with the metallic nature of LuNb$_6$Sn$_6$~\cite{ortiz2025stability}. A similar suppression of the DOS near $E_F$ has also been reported in other kagome materials, including CsV$_3$Sb$_5$~\cite{bhandari2024first}.

Additionally, we computed the FS of LuNb$_6$Sn$_6$ in its pristine (non-CDW) phase. Figure~\ref{Fig8}(c) shows the band-resolved FS obtained from DFT calculations. Two bands crossing the $E_F$ contribute to the FS, corresponding to the DFT band indices 93 and 94. The FS associated with band 94 consists of a large warped ellipsoidal pocket centered at the $\Gamma$ point together with a small pocket near the Brillouin zone boundary, whereas the FS associated with band 93 exhibits a chain-like sheet extending along the Brillouin zone boundary.

We computed all possible theoretical quantum oscillation frequencies from the FS sheets of both bands using the SKEAF code~\cite{julian2012numerical}, and the results are presented in Fig.~\ref{Fig10}. The experimentally observed dHvA frequencies are included for comparison. As shown in Fig.~\ref{Fig10}(a), all theoretical frequencies derived from Band 93 are below 500 T. At $\theta = 0^\circ$, the calculated frequency is close to the experimentally observed $F_\beta$. However, the deviation starts at higher $\theta$ values: the experimental $F_\beta$ decreases with angle, whereas the corresponding theoretical frequency exhibits an increasing trend. To examine the sensitivity of the calculated frequencies to the chemical potential, we manually shifted the $E_F$ by $\pm$20 meV, and the corresponding results are shown in Figs.~\ref{Fig10}(b) and \ref{Fig10}(c). Such a small change in the chemical potential can arise from slight deviations in stoichiometry, point defects, impurities, or doping, which modify the carrier concentration and consequently shift the $E_F$.

Shifting the $E_F$ downward further increases the discrepancy with the experimental $F_\beta$. In contrast, shifting the $E_F$ upward by 20 meV improves the agreement in the frequency magnitude up to approximately $\theta = 30^\circ$. Nevertheless, the angular dependence remains inconsistent with the experiment, as the calculated frequency continues to increase with $\theta$, while $F_\beta$ decreases. Since the angular dependence of the dHvA frequency directly reflects the geometry of the corresponding FS sheet~\cite{ando2013topological}, this discrepancy suggests that the FS geometry is modified by the CDW distortion. The theoretical frequencies derived from Band 94, shown in Fig.~\ref{Fig10}(d), are located near 85 T, 4000 T, and 7000 T. None of these frequencies are observed experimentally. As seen in the figure, the $E_F$ shift has a significant effect on the lowest frequency near 85 T but has little to no effect on the higher frequencies near 4000 T and 7000 T. 

Overall, the calculated frequencies for the pristine phase of LuNb$_6$Sn$_6$ cannot fully account for the experimentally observed dHvA results, while several frequencies predicted by DFT are absent in the measurements. These discrepancies indicate that the CDW distortion may reconstruct the FS in LuNb$_6$Sn$_6$, consistent with previous reports on several kagome materials~\cite{ortiz2021fermi,shrestha2023electronic,zhang2022emergence}. The FS geometry of this compound is expected to change upon entering the CDW phase because of the formation of a $\sqrt{3}\times\sqrt{3}\times 3$ supercell and the associated reconstruction of the electronic structure. However, explicit electronic-structure calculations for the CDW phase require large supercells and are computationally expensive, placing them beyond our current capability. A detailed investigation of the CDW-induced FS reconstruction is therefore beyond the scope of the present work and is left for future studies.

Understanding the FS of LuNb$_6$Sn$_6$ is important for clarifying how CDW order affects the low-energy electronic structure of HfFe$_6$Ge$_6$-type kagome metals. Within this broad family, LuNb$_6$Sn$_6$ and ScV$_6$Sn$_6$ occupy a special position, as they are the only known nonmagnetic members that display a CDW-like instability~\cite{ortiz2025stability, arachchige2022charge}. Previous quantum oscillation measurements on ScV$_6$Sn$_6$ showed that its CDW state coexists with small FS pockets, light cyclotron masses, and a Dirac-derived band with a nontrivial Berry phase~\cite{yi2024quantum,zheng2024quantum,shrestha2023electronic}. Our dHvA measurements reveal that LuNb$_6$Sn$_6$ has a comparatively simple low-energy FS, dominated by two small ellipsoidal pockets with light effective masses. This suggests that light Dirac-like quasiparticles can remain present in the CDW state, while the difference between LuNb$_6$Sn$_6$ and ScV$_6$Sn$_6$ points to a material-dependent reconstruction of the electronic structure.

\section{Summary}
We have investigated the properties of the newly discovered kagome metal LuNb$_6$Sn$_6$ using torque magnetometry, magnetization, and heat-capacity measurements. Temperature-dependent torque data reveal a clear anomaly at $T_{\mathrm{CDW}} = 85$ K associated with a CDW transition, which is further confirmed by heat-capacity measurements. Thermal hysteresis observed between cooling and warming curves in both torque and heat-capacity data establishes the first-order nature of the transition.

Torque and magnetization measurements in magnetic fields up to 14~T reveal pronounced dHvA oscillations with two dominant frequencies, $F_\alpha \approx 20$~T and $F_\beta \approx 200$~T, for magnetic fields applied along the $ab$ plane. Importantly, the frequencies observed in both torque and magnetization measurements are consistent across three samples, confirming their intrinsic origin in LuNb$_6$Sn$_6$. Angle-dependent measurements further show that both frequencies persist over a wide angle range and follow a behavior consistent with ellipsoidal FS pockets. 

The temperature dependence of the dHvA oscillation amplitudes was analyzed using the Lifshitz--Kosevich formalism to extract the effective mass, Dingle temperature, scattering time, mean free path, and quantum mobility for the observed orbits in all three samples. The extracted parameters, summarized in Table I, reveal anisotropic charge-carrier dynamics between the in-plane and out-of-plane directions, with enhanced carrier mobility along the out-of-plane direction. 

Additionally, our Landau fan diagram analysis reveals that the $F_{\alpha}$ pocket exhibits a nontrivial Berry phase, ($\Phi_B$), indicating its topological character. In contrast, the phase extracted for the $F_{\beta}$ pocket is consistent with a trivial Berry phase. However, because of the ambiguity associated with determining the phase shift factor ($\delta$), further studies are required to establish conclusively whether this pocket is topologically trivial. Overall, these results provide evidence for nontrivial electronic topology in LuNb$_6$Sn$_6$ and motivate further investigations of possible topological signatures in its electronic properties.

First-principles calculations of the electronic structure were performed to support the quantum oscillation results. The calculations reveal the presence of DP, FB, and VHSs near the $E_F$. Our calculations based on the pristine phase yield quantum oscillation frequencies that cannot fully reproduce the experimentally observed frequencies, suggesting that CDW-induced FS reconstruction plays a crucial role in determining the electronic structure of LuNb$_6$Sn$_6$. The detailed dHvA results, together with first-principles calculations, provide important insights into the FS geometry and electronic structure of LuNb$_6$Sn$_6$, and establish a foundation for future investigations of the CDW state and its effect on the electronic properties of this kagome material.

\section*{acknowledgements}
This work was supported in part by the U.S. Department of Energy, Office of Science, Office of Workforce Development for Teachers and Scientists (WDTS) under the Visiting Faculty Program (VFP) at Los Alamos National Laboratory, administered by the Oak Ridge Institute for Science and Education. The work at West Texas A\&M University (WTAMU) is supported by the Killgore Undergraduate and Graduate Student Research Grants, the Welch Foundation (Grant No. AE-0025) and the National Science Foundation (Award No. 2336011). B.R.O and J.Y. (sample discovery, single-crystal synthesis), were primarily supported by the U.S. Department of Energy (DOE), Office of Science, Basic Energy Sciences (BES), Materials Sciences and Engineering Division. Development of growth technique for enhanced-$T_{\mathrm{CDW}}$ LuNb$_6$Sn$_6$ was supported by the Laboratory Directed Research and Development Program of Oak Ridge National Laboratory, managed by UT-Battelle, LLC, for the US Department of Energy. C.A.M. acknowledges the support of the Laboratory Directed Research and Development program of Los Alamos National Laboratory under project number 20240225ER. A portion of this work was performed at the NHMFL, which is supported by National Science Foundation Cooperative Agreement No. DMR-2128556, the State of Florida, and the U.S. Department of Energy.

\section*{Data Availability}
The data that support the findings of this article are not publicly available. The data are available from the authors upon reasonable request.

\newpage

\bibliography{Bibliography_fixed_arxiv}

@article{shtefiienko2025electronic,
  title={{Electronic structure of YV$_6$Sn$_6$ probed by de Haas--van Alphen oscillations and density functional theory}},
  author={Shtefiienko, Kyryl and Phillips, Cole and Mozaffari, Shirin and Madhogaria, Richa P and Meier, William R and Mandrus, David G and Graf, David E and Shrestha, Keshav},
  journal={APL Quantum},
  volume={2},
  pages={016118},
  year={2025},
  publisher={AIP Publishing}
}

@article{yin2022topological,
  title = {{Topological kagome magnets and superconductors}},
  author={Yin, Jia-Xin and Lian, Biao and Hasan, M Zahid},
  journal={Nature},
  volume={612},
  number={7941},
  pages={647--657},
  year={2022},
  publisher={Nature Publishing Group UK London}
}

@article{wilson2024v3sb5,
  title = {{AV$_3$Sb$_5$ kagome superconductors}},
  author={Wilson, Stephen D and Ortiz, Brenden R},
  journal={ Nat. Rev. Mater.},
  volume={612},
  pages={420},
  year={2024},
  publisher={Nature Publishing Group UK London}
}

@article{wang2023quantum,
  title = {{Quantum states and intertwining phases in kagome materials}},
  author={Wang, Yaojia and Wu, Heng and McCandless, Gregory T and Chan, Julia Y and Ali, Mazhar N},
  journal = {Nat. Rev. Phys.},
  volume={5},
  number={11},
  pages={635--658},
  year={2023},
  publisher={Nature Publishing Group UK London}
}

@article{mekata2003kagome,
  title = {Kagome: The story of the basketweave lattice},
  author={Mekata, Mamoru},
  journal={Phys. Today},
  volume={56},
  number={2},
  pages={12--13},
  year={2003},
  publisher={AIP Publishing}
}

@article{peng2021realizing,
  title = {{Realizing kagome band structure in two-dimensional kagome surface states of RV$_6$Sn$_6$ (R = Gd, Ho)}},
  author={Peng, Shuting and Han, Yulei and Pokharel, Ganesh and Shen, Jianchang and Li, Zeyu and Hashimoto, Makoto and Lu, Donghui and Ortiz, Brenden R and Luo, Yang and Li, Houchen and others},
  journal={Phys. Rev. Lett.},
  volume={127},
  number={26},
  pages={266401},
  year={2021},
  publisher={APS}
}

@article{bhandari2025pressure,
  title = {{Pressure effects on the electronic structure of the kagome metal CsV$_3$Sb$_5$}},
  author={Bhandari, Shalika R and Gusain, Vivek and Zeeshan, Mohd and Rai, DP and Shrestha, Keshav},
  journal={APL Quantum},
  volume={2},
  pages={016133},
  year={2025},
  publisher={AIP Publishing}
}

@article{shrestha2023electronic,
  title = {{Electronic properties of kagome metal ScV$_6$Sn$_6$ using high-field torque magnetometry}},
  author={Shrestha, Keshav and Regmi, Binod and Pokharel, Ganesh and Kim, Seong-Gon and Wilson, Stephen D and Graf, David E and Magar, Birendra A and Phillips, Cole and Nguyen, Thinh},
  journal={Phys. Rev. B},
  volume={108},
  number={24},
  pages={245119},
  year={2023},
  publisher={APS}
}

@article{arachchige2022charge,
  title = {{Charge density wave in kagome lattice intermetallic ScV$_6$Sn$_6$}},
  author={Arachchige, Hasitha W Suriya and Meier, William R and Marshall, Madalynn and Matsuoka, Takahiro and Xue, Rui and McGuire, Michael A and Hermann, Raphael P and Cao, Huibo and Mandrus, David},
  journal={Phys. Rev. Lett.},
  volume={129},
  number={21},
  pages={216402},
  year={2022},
  publisher={APS}
}

@article{werhahn2022kagome,
  title = {{The kagome metals CsTi$_3$Bi$_5$ and RbTi$_3$Bi$_5$}},
  author = {Werhahn, Dominik and Ortiz, Brenden R. and Hay, Aurland K. and Wilson, Stephen D. and Seshadri, Ram and Johrendt, Dirk},
  journal = {Z. Naturforsch. B},
  volume = {77},
  number = {11--12},
  pages = {757--764},
  year = {2022},
  publisher = {De Gruyter}
}

@article{ortiz2023evolution,
  title = {{Evolution of Highly Anisotropic Magnetism in the Titanium-Based Kagome Metals LnTi$_3$Bi$_4$ (Ln: La{\textperiodcentered}{\textperiodcentered}{\textperiodcentered} Gd$^{3+}$, Eu$^{2+}$, Yb$^{2+}$)}},
  author = {Ortiz, Brenden R. and Miao, Hu and Parker, David S. and Yang, Fazhi and Samolyuk, German D. and Clements, Eleanor M. and Rajapitamahuni, Anil and Yilmaz, Turgut and Vescovo, Elio and Yan, Jiaqiang and others},
  journal = {Chem. Mater.},
  volume = {35},
  number = {22},
  pages = {9756--9773},
  year = {2023},
  publisher = {ACS Publications}
}

@article{ortiz2024intricate,
  title = {{Intricate Magnetic Landscape in Antiferromagnetic Kagome Metal TbTi$_3$Bi$_4$ and Interplay with Ln$_{2-x}$Ti$_{6+x}$Bi$_9$ (Ln: Tb{\textperiodcentered}{\textperiodcentered}{\textperiodcentered} Lu) Shurikagome Metals}},
  author = {Ortiz, Brenden R. and Zhang, Heda and G{\'o}rnicka, Karolina and Parker, David S. and Samolyuk, German D. and Yang, Fazhi and Miao, Hu and Lu, Qiangsheng and Moore, Robert G. and May, Andrew F. and others},
  journal = {Chem. Mater.},
  volume = {36},
  number = {16},
  pages = {8002--8014},
  year = {2024},
  publisher = {ACS Publications}
}

@article{phillips2025electrical,
  title = {{Electrical transport and torque magnetometry studies of the kagome compound LuV$_6$Sn$_6$ under high magnetic fields}},
  author = {Phillips, Cole and Shtefiienko, Kyryl and Mozaffari, Shirin and Madhogaria, Richa P. and Meier, William R. and Savvidou, Aikaterini Flessa and Casas, Brian W. and Balicas, Luis and Mandrus, David G. and Graf, David E. and others},
  journal = {Phys. Rev. B},
  volume = {111},
  number = {15},
  pages = {155136},
  year = {2025},
  publisher = {APS}
}

@article{ye2018massive,
  title = {{Massive Dirac fermions in a ferromagnetic kagome metal}},
  author={Ye, Linda and Kang, Mingu and Liu, Junwei and Von Cube, Felix and Wicker, Christina R and Suzuki, Takehito and Jozwiak, Chris and Bostwick, Aaron and Rotenberg, Eli and Bell, David C and others},
  journal={Nature},
  volume={555},
  number={7698},
  pages={638--642},
  year={2018},
  publisher={Nature Publishing Group UK London}
}

@article{liu2018giant,
  title = {{Giant anomalous Hall effect in a ferromagnetic kagome-lattice semimetal}},
  author={Liu, Enke and Sun, Yan and Kumar, Nitesh and Muechler, Lukas and Sun, Aili and Jiao, Lin and Yang, Shuo-Ying and Liu, Defa and Liang, Aiji and Xu, Qiunan and others},
  journal={Nat. Phys.},
  volume={14},
  number={11},
  pages={1125--1131},
  year={2018},
  publisher={Nature Publishing Group UK London}
}

@article{li2025giant,
  title = {{Giant anomalous Hall effect in the kagome nodal surface semimetal Fe$_3$Ge}},
  author = {Li, Shu-Xiang and Wang, Wencheng and Xu, Sheng and Li, Tianhao and Li, Zheng and Wang, Jinjin and Mi, Jun-Jian and Tao, Qian and Tang, Feng and Wan, Xiangang and others},
  journal = {Phys. Rev. B},
  volume = {112},
  number = {18},
  pages = {184418},
  year = {2025},
  publisher = {APS}
}

@article{hou2023effect,
  title = {{Effect of hydrostatic pressure on the unconventional charge density wave and superconducting properties in two distinct phases of doped kagome superconductors CsV$_{3-x}$Ti$_x$Sb$_{5}$}},
  author={Hou, J and Chen, KY and Sun, JP and Zhao, Z and Zhang, YH and Shan, PF and Wang, NN and Zhang, H and Zhu, K and Uwatoko, Y and others},
  journal={Phys. Rev. B},
  volume={107},
  number={14},
  pages={144502},
  year={2023},
  publisher={APS}
}

@article{yang2022titanium,
  title = {{Titanium doped kagome superconductor CsV$_{3-x}$Ti$_x$Sb$_{5}$ and two distinct phases}},
  author={Yang, Haitao and Huang, Zihao and Zhang, Yuhang and Zhao, Zhen and Shi, Jinan and Luo, Hailan and Zhao, Lin and Qian, Guojian and Tan, Hengxin and Hu, Bin and others},
  journal={Sci. Bull.},
  volume={67},
  number={21},
  pages={2176--2185},
  year={2022},
  publisher={Elsevier}
}

@article{oey2022fermi,
  title = {{Fermi level tuning and double-dome superconductivity in the kagome metal CsV$_3$Sb$_{5-x}$Sn$_x$}},
  author={Oey, Yuzki M and Ortiz, Brenden R and Kaboudvand, Farnaz and Frassineti, Jonathan and Garcia, Erick and Cong, Rong and Sanna, Samuele and Mitrovi{\'c}, Vesna F and Seshadri, Ram and Wilson, Stephen D},
  journal = {Phys. Rev. Mater.},
  volume={6},
  number={4},
  pages={L041801},
  year={2022},
  publisher={APS}
}

@article{kautzsch2023incommensurate,
  title = {{Incommensurate charge-stripe correlations in the kagome superconductor CsV$_3$Sb$_{5-x}$Sn$_x$}},
  author={Kautzsch, Linus and Oey, Yuzki M and Li, Hong and Ren, Zheng and Ortiz, Brenden R and Pokharel, Ganesh and Seshadri, Ram and Ruff, Jacob and Kongruengkit, Terawit and Harter, John W and others},
  journal = {npj Quantum Mater.},
  volume={8},
  number={1},
  pages={37},
  year={2023},
  publisher={Nature Publishing Group UK London}
}

@article{qian2021revealing,
  title = {{Revealing the competition between charge density wave and superconductivity in CsV$_3$Sb$_5$ through uniaxial strain}},
  author={Qian, Tiema and Christensen, Morten H and Hu, Chaowei and Saha, Amartyajyoti and Andersen, Brian M and Fernandes, Rafael M and Birol, Turan and Ni, Ni},
  journal={Phys. Rev. B},
  volume={104},
  number={14},
  pages={144506},
  year={2021},
  publisher={APS}
}

@article{lin2024uniaxial,
  title = {{Uniaxial strain tuning of charge modulation and singularity in a kagome superconductor}},
  author={Lin, Chun and Consiglio, Armando and Forslund, Ola Kenji and K{\"u}spert, Julia and Denner, M Michael and Lei, Hechang and Louat, Alex and Watson, Matthew D and Kim, Timur K and Cacho, Cephise and others},
  journal={Nat. Commun.},
  volume={15},
  number={1},
  pages={10466},
  year={2024},
  publisher={Nature Publishing Group UK London}
}

@article{yang2023plane,
  title = {{In-plane uniaxial-strain tuning of superconductivity and charge-density wave in  CsV$_3$Sb$_5$}},
  author={Yang, Xiaoran and Tang, Qi and Zhou, Qiuyun and Wang, Huaiping and Li, Yi and Fu, Xue and Zhang, Jiawen and Song, Yu and Yuan, Huiqiu and Dai, Pengcheng and others},
  journal={Chin. Phys. B},
  volume={32},
  number={12},
  pages={127101},
  year={2023},
  publisher={Chinese Physical Society and IOP Publishing Ltd}
}

@article{yu2021unusual,
  title = {{Unusual competition of superconductivity and charge-density-wave state in a compressed topological kagome metal}},
  author={Yu, FH and Ma, DH and Zhuo, WZ and Liu, SQ and Wen, XK and Lei, Bin and Ying, JJ and Chen, XH},
  journal={Nat. Commun.},
  volume={12},
  number={1},
  pages={3645},
  year={2021},
  publisher={Nature Publishing Group UK London}
}

@article{chen2021double,
  title = {{Double superconducting dome and triple enhancement of T$_c$ in the kagome superconductor CsV$_3$Sb$_5$ under high pressure}},
  author={Chen, KY and Wang, NN and Yin, QW and Gu, YH and Jiang, K and Tu, ZJ and Gong, CS and Uwatoko, Y and Sun, JP and Lei, HC and others},
  journal={Phys. Rev. Lett.},
  volume={126},
  number={24},
  pages={247001},
  year={2021},
  publisher={APS}
}

@article{phillips2024fermi,
  title = {{Fermi surface reconstruction under pressure in the kagome metal  CsV$_3$Sb$_5$}},
  author={Phillips, Cole and Shtefiienko, Kyryl and Nguyen, Thinh and Capa Salinas, Andrea N and Magar, Birendra A and Pokharel, Ganesh and Wilson, Stephen D and Graf, David E and Shrestha, Keshav},
  journal={Phys. Rev. B},
  volume={110},
  number={20},
  pages={205135},
  year={2024},
  publisher={APS}
}

@article{di2026kagome,
  title = {{Kagome metals}},
  author={Di Sante, Domenico and Neupert, Titus and Sangiovanni, Giorgio and Thomale, Ronny and Comin, Riccardo and Checkelsky, Joseph G and Zeljkovic, Ilija and Wilson, Stephen D},
  journal={Rev. Mod. Phys.},
  volume={98},
  number={1},
  pages={015002},
  year={2026},
  publisher={APS}
}

@article{hu2023electronic,
  title = {{Electronic landscape of kagome superconductors AV$_3$Sb$_5$ (A = K, Rb, Cs) from angle-resolved photoemission spectroscopy}},
  author={Hu, Yong and Wu, Xianxin and Schnyder, Andreas P and Shi, Ming},
  journal={npj Quantum Mater.},
  volume={8},
  number={1},
  pages={67},
  year={2023},
  publisher={Nature Publishing Group UK London}
}

@article{wang2024topological,
  title = {{Topological quantum materials with kagome lattice}},
  author={Wang, Qi and Lei, Hechang and Qi, Yanpeng and Felser, Claudia},
  journal={Acc. Mater. Res.},
  volume={5},
  number={7},
  pages={786--796},
  year={2024},
  publisher={ACS Publications}
}

@article{jiang2023kagome,
  title = {{Kagome superconductors AV$_3$Sb$_5$ (A = K, Rb, Cs)}},
  author={Jiang, Kun and Wu, Tao and Yin, Jia-Xin and Wang, Zhenyu and Hasan, M Zahid and Wilson, Stephen D and Chen, Xianhui and Hu, Jiangping},
  journal = {Natl. Sci. Rev.},
  volume={10},
  number={2},
  pages={nwac199},
  year={2023},
  publisher={Oxford University Press}
}

@article{bernevig2022progress,
  title = {{Progress and prospects in magnetic topological materials}},
  author={Bernevig, B Andrei and Felser, Claudia and Beidenkopf, Haim},
  journal={Nature},
  volume={603},
  number={7899},
  pages={41--51},
  year={2022},
  publisher={Nature Publishing Group UK London}
}

@article{yu2021concurrence,
  title = {{Concurrence of anomalous Hall effect and charge density wave in a superconducting topological kagome metal}},
  author={Yu, FH and Wu, T and Wang, ZY and Lei, B and Zhuo, WZ and Ying, JJ and Chen, XH},
  journal={Phys. Rev. B},
  volume={104},
  number={4},
  pages={L041103},
  year={2021},
  publisher={APS}
}

@article{yin2021superconductivity,
  title = {{Superconductivity and normal-state properties of kagome metal RbV$_3$Sb$_5$ single crystals}},
  author={Yin, Qiangwei and Tu, Zhijun and Gong, Chunsheng and Fu, Yang and Yan, Shaohua and Lei, Hechang},
  journal={Chin. Phys. Lett.},
  volume={38},
  number={3},
  pages={037403},
  year={2021},
  publisher={IOP Publishing}
}

@article{shrestha2023high,
  title = {{High quantum oscillation frequencies and nontrivial topology in kagome superconductor KV$_3$Sb$_5$ probed by torque magnetometry up to 45 T}},
  author={Shrestha, K and Shi, M and Regmi, B and Nguyen, T and Miertschin, D and Fan, K and Deng, LZ and Aryal, N and Kim, S-G and Graf, DE and others},
  journal={Phys. Rev. B},
  volume={107},
  number={15},
  pages={155128},
  year={2023},
  publisher={APS}
}

@article{yang2020giant,
  title = {{Giant, unconventional anomalous Hall effect in the metallic frustrated magnet candidate, KV$_3$Sb$_5$}},
  author={Yang, Shuo-Ying and Wang, Yaojia and Ortiz, Brenden R and Liu, Defa and Gayles, Jacob and Derunova, Elena and Gonzalez-Hernandez, Rafael and {\v{S}}mejkal, Libor and Chen, Yulin and Parkin, Stuart SP and others},
  journal={Sci. Adv.},
  volume={6},
  number={31},
  pages={eabb6003},
  year={2020},
  publisher={American Association for the Advancement of Science}
}

@article{nakayama2022carrier,
  title={{Carrier injection and manipulation of charge-density wave in kagome superconductor CsV$_3$Sb$_5$}},
  author={Nakayama, Kosuke and Li, Yongkai and Kato, Takemi and Liu, Min and Wang, Zhiwei and Takahashi, Takashi and Yao, Yugui and Sato, Takafumi},
  journal={Phys. Rev. X},
  volume={12},
  number={1},
  pages={011001},
  year={2022},
  publisher={APS}
}

@article{ortiz2021fermi,
  title = {{Fermi surface mapping and the nature of charge-density-wave order in the kagome superconductor CsV$_3$Sb$_5$}},
  author={Ortiz, Brenden R and Teicher, Samuel ML and Kautzsch, Linus and Sarte, Paul M and Ratcliff, Noah and Harter, John and Ruff, Jacob PC and Seshadri, Ram and Wilson, Stephen D},
  journal={Phys. Rev. X},
  volume={11},
  number={4},
  pages={041030},
  year={2021},
  publisher={APS}
}

@article{shrestha2023fermi,
  title = {{Fermi surface mapping of the kagome superconductor RbV$_3$Sb$_5$ using de Haas-van Alphen oscillations}},
  author={Shrestha, Keshav and Shi, Mengzhu and Nguyen, Thinh and Miertschin, Duncan and Fan, Kaibao and Deng, Liangzi and Graf, David E and Chen, Xianhui and Chu, Ching-Wu},
  journal={Phys. Rev. B},
  volume={107},
  number={7},
  pages={075120},
  year={2023},
  publisher={APS}
}

@article{luo2022electronic,
  title={{Electronic nature of charge density wave and electron-phonon coupling in kagome superconductor KV$_3$Sb$_5$}},
  author={Luo, Hailan and Gao, Qiang and Liu, Hongxiong and Gu, Yuhao and Wu, Dingsong and Yi, Changjiang and Jia, Junjie and Wu, Shilong and Luo, Xiangyu and Xu, Yu and others},
  journal={Nat. Commun.},
  volume={13},
  number={1},
  pages={273},
  year={2022},
  publisher={Nature Publishing Group UK London}
}

@article{fu2021quantum,
  title = {{Quantum transport evidence of topological band structures of kagome superconductor CsV$_3$Sb$_5$}},
  author={Fu, Yang and Zhao, Ningning and Chen, Zheng and Yin, Qiangwei and Tu, Zhijun and Gong, Chunsheng and Xi, Chuanying and Zhu, Xiangde and Sun, Yuping and Liu, Kai and others},
  journal={Phys. Rev. Lett.},
  volume={127},
  number={20},
  pages={207002},
  year={2021},
  publisher={APS}
}

@article{zhang2022emergence,
  title = {{Emergence of large quantum oscillation frequencies in thin flakes of the kagome superconductor CsV$_3$Sb$_5$}},
  author={Zhang, W and Wang, Lingfei and Tsang, Chun Wai and Liu, Xinyou and Xie, Jianyu and Yu, Wing Chi and Lai, Kwing To and Goh, Swee K},
  journal={Phys. Rev. B},
  volume={106},
  number={19},
  pages={195103},
  year={2022},
  publisher={APS}
}

@article{broyles2022effect,
  title = {{Effect of the interlayer ordering on the Fermi surface of Kagome superconductor CsV$_3$Sb$_5$ revealed by quantum oscillations}},
  author={Broyles, Christopher and Graf, David and Yang, Haitao and Dong, Xiaoli and Gao, Hongjun and Ran, Sheng},
  journal={Phys. Rev. Lett.},
  volume={129},
  number={15},
  pages={157001},
  year={2022},
  publisher={APS}
}

@article{shrestha2022nontrivial,
  title = {{Nontrivial Fermi surface topology of the kagome superconductor CsV$_3$Sb$_5$ probed by de Haas--van Alphen oscillations}},
  author={Shrestha, K and Chapai, R and Pokharel, Bal K and Miertschin, D and Nguyen, T and Zhou, X and Chung, DY and Kanatzidis, MG and Mitchell, JF and Welp, U and others},
  journal={Phys. Rev. B},
  volume={105},
  number={2},
  pages={024508},
  year={2022},
  publisher={APS}
}

@article{rosenberg2024probing,
  title = {{Probing the van Hove singularity of the kagome metal YV$_6$Sn$_6$ through quantum oscillations}},
  author={Rosenberg, Elliott and DeStefano, Jonathan M and Lee, Yongbin and Hu, Chaowei and Shi, Yue and Graf, David and Benjamin, Shermane M and Ke, Liqin and Chu, Jiun-Haw},
  journal={Phys. Rev. B},
  volume={110},
  number={3},
  pages={035119},
  year={2024},
  publisher={APS}
}

@article{zheng2024quantum,
  title={{Quantum oscillations evidence for topological bands in kagome metal ScV$_6$Sn$_6$}},
  author={Zheng, Guoxin and Zhu, Yuan and Mozaffari, Shirin and Mao, Ning and Chen, Kuan-Wen and Jenkins, Kaila and Zhang, Dechen and Chan, Aaron and Arachchige, Hasitha W Suriya and Madhogaria, Richa P and others},
  journal={J. Phys.: Condens. Matter},
  volume={36},
  number={21},
  pages={215501},
  year={2024},
  publisher={IOP Publishing}
}

@article{yi2024quantum,
  title={{Quantum oscillations revealing topological band in kagome metal ScV$_6$Sn$_6$}},
  author={Yi, Changjiang and Feng, Xiaolong and Mao, Ning and Yanda, Premakumar and Roychowdhury, Subhajit and Zhang, Yang and Felser, Claudia and Shekhar, Chandra},
  journal={Phys. Rev. B},
  volume={109},
  number={3},
  pages={035124},
  year={2024},
  publisher={APS}
}

@article{ortiz2019new,
  title = {{New kagome prototype materials: discovery of KV$_3$Sb$_5$, RbV$_3$Sb$_5$, and CsV$_3$Sb$_5$}},
  author={Ortiz, Brenden R and Gomes, L{\'\i}dia C and Morey, Jennifer R and Winiarski, Michal and Bordelon, Mitchell and Mangum, John S and Oswald, Iain WH and Rodriguez-Rivera, Jose A and Neilson, James R and Wilson, Stephen D and others},
  journal={Phys. Rev. Mater.},
  volume={3},
  number={9},
  pages={094407},
  year={2019},
  publisher={APS}
}

@article{ortiz2020cs,
  title = {{CsV$_3$Sb$_5$: A Z$_2$ topological kagome metal with a superconducting ground state}},
  author={Ortiz, Brenden R and Teicher, Samuel ML and Hu, Yong and Zuo, Julia L and Sarte, Paul M and Schueller, Emily C and Abeykoon, AM Milinda and Krogstad, Matthew J and Rosenkranz, Stephan and Osborn, Raymond and others},
  journal={Phys. Rev. Lett.},
  volume={125},
  number={24},
  pages={247002},
  year={2020},
  publisher={APS}
}

@article{ortiz2025stability,
  title = {{Stability Frontiers in the AM$_6$X$_6$ Kagome Metals: The LnNb$_6$Sn$_6$ (Ln: Ce--Lu, Y) Family and Density-Wave Transition in LuNb$_6$Sn$_6$}},
  author = {Ortiz, Brenden R and Meier, William R and Pokharel, Ganesh and Chamorro, Juan and Yang, Fazhi and Mozaffari, Shirin and Thaler, Alex and Gomez Alvarado, Steven J and Zhang, Heda and Parker, David S and others},
  journal = {J. Am. Chem. Soc.},
  volume = {147},
  number = {6},
  pages = {5279--5292},
  year = {2025},
  publisher = {ACS Publications}
}

@article{lou2025orbital,
  title = {{Orbital-selective band modifications in a charge-ordered kagome metal LuNb$_6$ Sn$_6$}},
  author={Lou, Rui and Zhang, Yumeng and Cheng, Erjian and Feng, Xiaolong and Fedorov, Alexander and Li, Zongkai and Luo, Yixuan and Generalov, Alexander and Ma, Haiyang and Wei, Quanxing and others},
  journal={arXiv preprint arXiv:2504.04019},
  year={2025},
}

@article{ingham2024theory,
  title = {{Theory of excitonic order in kagome metals ScV$_6$Sn$_6$ and LuNb$_6$Sn$_6$}},
  author={Ingham, Julian and Consiglio, Armando and di Sante, Domenico and Thomale, Ronny and Scammell, Harley D},
  journal={arXiv preprint arXiv:2410.16365},
  year={2024},
}

@article{yang2025fermi,
  title={{Fermi-surface driven frustrations in charge ordered kagome metal LuNb$_6$ Sn$_6$}},
  author={Yang, FZ and Huang, X and Tan, Hengxin and Kundu, A and Kim, S and Thinel, M and Ingham, J and Rajapitamahuni, Anil and Cai, YQ and Nelson, C and others},
  journal={Phys. Rev. B},
  volume={112},
  pages={245113},
  year={2025},
}

@article{meier2025pressure,
  title = {{Pressure suppresses the density wave order in kagome metal LuNb$_6$ Sn$_6$}},
  author={Meier, William R and Graf, David E and Ortiz, Brenden R and Mozaffari, Shirin and Mandrus, David},
  journal={Phys. Rev. Mater.},
  volume={9},
  number={8},
  pages={L082001},
  year={2025},
  publisher={APS}
}

@article{yan2017flux,
  title={{Flux growth in a horizontal configuration: An analog to vapor transport growth}},
  author={Yan, J-Q and Sales, Brian C and Susner, Michael A and Mcguire, Michael A},
  journal={Phys. Rev. Mater.},
  volume={1},
  number={2},
  pages={023402},
  year={2017},
  publisher={APS}
}

@article{QE-2017,
author={Giannozzi, Paolo and Andreussi, Oliviero and Brumme, Thomas and Bunau, Oana and Buongiorno Nardelli, M and Calandra, Matteo and Car, Roberto and Cavazzoni, Carlo and Ceresoli, Davide and Cococcioni, Matteo and others},
title={{Advanced capabilities for materials modelling with QUANTUM ESPRESSO}},
journal={J. Phys.: Condens. Matter},
volume={29},
number={46},
pages={465901},
year={2017},
}

@article{QE-2009,
author={Giannozzi, Paolo and Baroni, Stefano and Bonini, Nicola and Calandra, Matteo and Car, Roberto and Cavazzoni, Carlo and Ceresoli, Davide and Chiarotti, Guido L and Cococcioni, Matteo and Dabo, Ismaila and others},
Journal = {J. Phys.: Condens. Matter},
Number = {39},
Pages = {395502},
Title = {{QUANTUM ESPRESSO: a modular and open-source software project for quantum simulations of materials}},
Volume = {21},
Year = {2009}
}

@article{perdew1996generalized,
  title={{Generalized gradient approximation made simple}},
  author={Perdew, John P and Burke, Kieron and Ernzerhof, Matthias},
  journal={Phys. Rev. Lett.},
  volume={77},
  number={18},
  pages={3865},
  year={1996},
  publisher={APS}
}

@article{perdew2008restoring,
  title={{Restoring the density-gradient expansion for exchange in solids and surfaces}},
  author={Perdew, John P and Ruzsinszky, Adrienn and Csonka, G{\'a}bor I and Vydrov, Oleg A and Scuseria, Gustavo E and Constantin, Lucian A and Zhou, Xiaolan and Burke, Kieron},
  journal={Phys. Rev. Lett.},
  volume={100},
  number={13},
  pages={136406},
  year={2008},
  publisher={APS}
}

@article{monkhorst1976special,
  title={{Special points for Brillouin-zone integrations}},
  author={Monkhorst, Hendrik J and Pack, James D},
  journal={Phys. Rev. B},
  volume={13},
  number={12},
  pages={5188},
  year={1976},
  publisher={APS}
}

@book{Shoenberg,
author = {Shoenberg, D.},
title = {{Magnetic {O}scillations in {M}etals}},
isbn = {9780521224802},
publisher = {Cambridge University Press},
year = {1984},
}

@article{ando2013topological,
  title = {Topological insulator materials},
  author={Ando, Yoichi},
  journal={J. Phys. Soc. Jpn.},
  volume={82},
  number={10},
  pages={102001},
  year={2013},
}

@article{shrestha2014shubnikov,
  title = {{Shubnikov-de Haas oscillations from topological surface states of metallic Bi$_{2}$Se$_{2.1}$Te$_{0.9}$}},
  author={Shrestha, Keshav and Marinova, Vera and Lorenz, Bernd and Chu, Paul CW},
  journal={Phys. Rev. B},
  volume={90},
  number={24},
  pages={241111},
  year={2014},
  publisher={APS}
}

@article{shrestha2018evidence,
  title={{Evidence of a 2D Fermi surface due to surface states in a p-type metallic Bi$_2$Te$_3$}},
  author={Shrestha, K and Marinova, V and Lorenz, B and Chu, CW},
  journal={J. Phys.: Condens. Matter},
  volume={30},
  number={18},
  pages={185601},
  year={2018},
  publisher={IOP Publishing}
}

@article{rehfuss2024quantum,
  title={{Quantum oscillations in kagome metals CsTi$_3$Bi$_5$ and RbTi$_3$Bi$_5$}},
  author={Rehfuss, Zackary and Broyles, Christopher and Graf, David and Li, Yongkang and Tan, Hengxin and Zhao, Zhen and Liu, Jiali and Zhang, Yuhang and Dong, Xiaoli and Yang, Haitao and others},
  journal={Phys. Rev. Mater.},
  volume={8},
  number={2},
  pages={024003},
  year={2024},
  publisher={APS}
}

@article{campbell2017quantum,
  title={{Quantum oscillations in the anomalous spin density wave state of FeAs}},
  author={Campbell, Daniel J and Eckberg, Chris and Wang, Kefeng and Wang, Limin and Hodovanets, Halyna and Graf, Dave and Parker, David and Paglione, Johnpierre},
  journal={Phys. Rev. B},
  volume={96},
  number={7},
  pages={075120},
  year={2017},
  publisher={APS}
}

@article{sebastian2008quantum,
  title={{Quantum oscillations in the parent magnetic phase of an iron arsenide high temperature superconductor}},
  author={Sebastian, Suchitra E and Gillett, J and Harrison, N and Lau, PHC and Singh, David J and Mielke, CH and Lonzarich, GG},
  journal={J. Phys.: Condens. Matter},
  volume={20},
  number={42},
  pages={422203},
  year={2008},
}

@article{dhital2024fermi,
  title = {{Fermi surface of the magnetic kagome compound GdV$_6$Sn$_6$ investigated using de Haas--van Alphen oscillations}},
  author = {Dhital, C. and Pokharel, G. and Wilson, B. and Kendrick, I. and Asmar, M. M. and Graf, D. and Guerrero-Sanchez, J. and Gonzalez-Hernandez, R. and Wilson, S. D.},
  journal = {Phys. Rev. B},
  volume = {109},
  number = {23},
  pages = {235145},
  year = {2024},
}

@article{prodan2024anisotropic,
  title = {{Anisotropic charge transport in the easy-plane kagome ferromagnet Fe$_3$Sn}},
  author={Prodan, Lilian and Chmeruk, Artem and Chioncel, Liviu and Tsurkan, Vladimir and K{\'e}zsm{\'a}rki, Istv{\'a}n},
  journal={Phys. Rev. B},
  volume={110},
  number={9},
  pages={094407},
  year={2024},
}

@article{ma2025anisotropic,
  title = {{Anisotropic transport properties and topological Hall effect in the annealed kagome antiferromagnet FeGe}},
  author={Ma, Jiajun and Shi, Chenfei and Cao, Yantao and Zhang, Yuwei and Li, Yazhou and Liao, Jiaxing and Wang, Jialu and Jiao, Wenhe and Guo, Hanjie and Xu, Chenchao and others},
  journal={Sci. China Phys. Mech. Astron.},
  volume={68},
  number={3},
  pages={237412},
  year={2025},
}

@article{hu2017nearly,
  title = {{Nearly massless Dirac fermions and strong Zeeman splitting in the nodal-line semimetal ZrSiS probed by de Haas--van Alphen quantum oscillations}},
  author={Hu, Jin and Tang, Zhijie and Liu, Jinyu and Zhu, Yanglin and Wei, Jiang and Mao, Zhiqiang},
  journal={Phys. Rev. B},
  volume={96},
  number={4},
  pages={045127},
  year={2017},
}

@article{miertschin2024anisotropic,
  title = {{Anisotropic quantum transport in ZrSiS probed by high-field torque magnetometry}},
  author = {Miertschin, Duncan and Nguyen, Thinh and Bhandari, Shalika R. and Shtefiienko, Kyryl and Phillips, Cole and Magar, Birendra A. and Sankar, Raman and Graf, David E. and Shrestha, Keshav},
  journal = {Phys. Rev. B},
  volume = {110},
  number = {8},
  pages = {085140},
  year = {2024},
}

@article{bhandari2024first,
  title = {{First-principles study of the electronic structure, Z$_2$ invariant, and quantum oscillation in the kagome material CsV$_3$Sb$_5$}},
  author={Bhandari, Shalika R and Zeeshan, Mohd and Gusain, Vivek and Shrestha, Keshav and Rai, DP},
  journal={APL Quantum},
  volume={1},
  pages={046118},
  year={2024},
}

@article{julian2012numerical,
  title = {{Numerical extraction of de Haas--van Alphen frequencies from calculated band energies}},
  author = {Julian, S. R.},
  journal = {Comput. Phys. Commun.},
  volume = {183},
  number = {2},
  pages = {324--332},
  year = {2012},
}

\newpage

\begin{figure}
  \centering
  \includegraphics[width=1.0\linewidth]{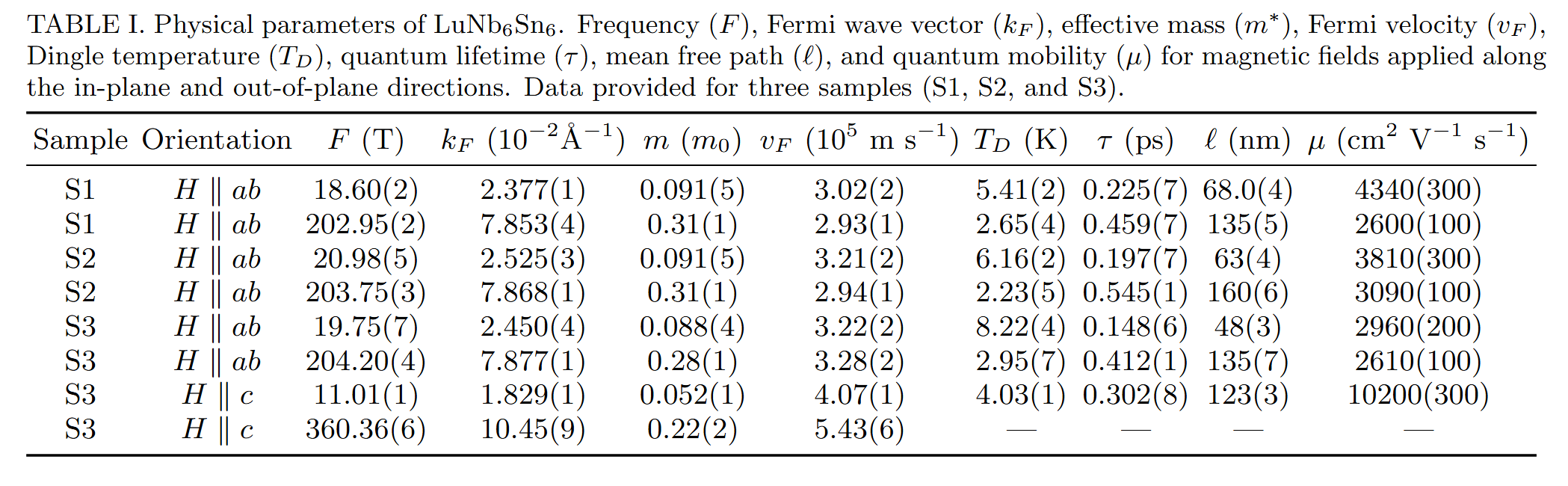}\label{table:parameters}
\end{figure}

\begin{figure}
  \centering
  \includegraphics[width=1.0\linewidth]{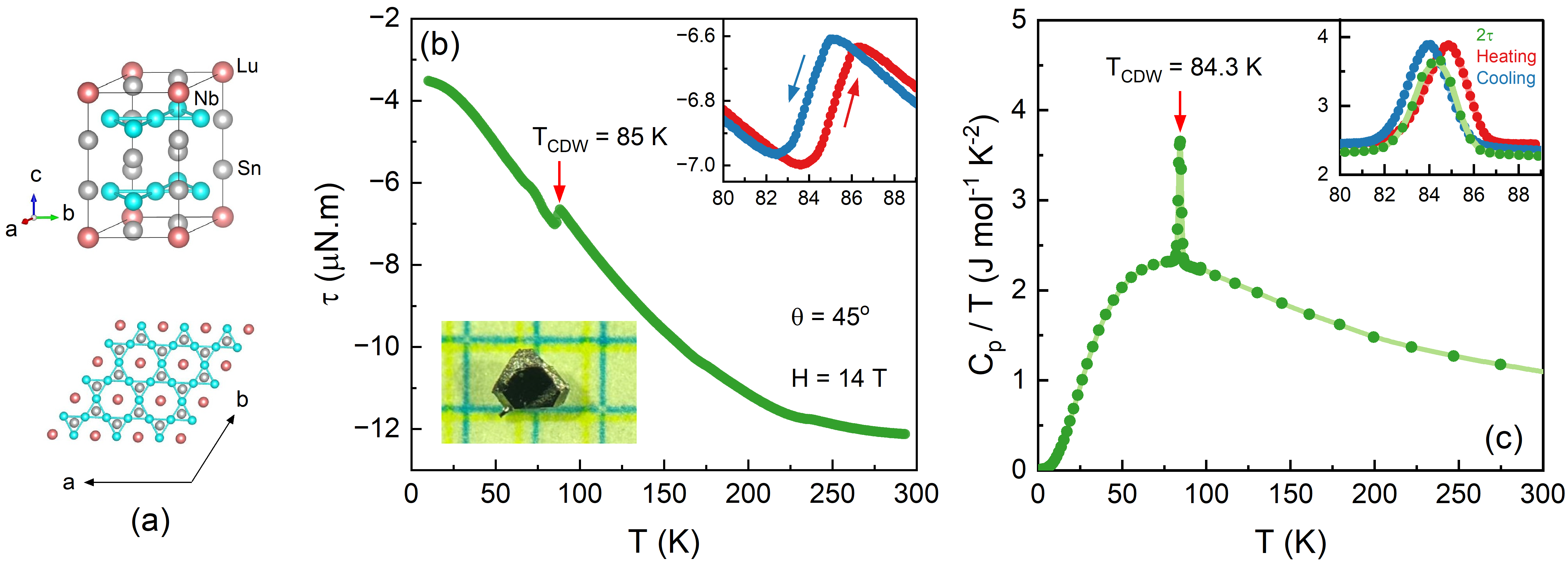}
  \caption{(a) Crystal structure of LuNb$_6$Sn$_6$: the unit cell (top) and top view (bottom), highlighting the Nb kagome network. (b) Temperature dependence of the magnetic torque at fixed field magnitude and direction for a LuNb$_6$Sn$_6$ single crystal (Sample S1). A clear anomaly at $T_{\mathrm{CDW}} = 85$ K (arrow) indicates the onset of CDW order. Upper inset: zoomed-in view near $T_{\mathrm{CDW}}$, showing the transition and hysteresis between cooling and heating curves. Lower inset: Optical image of one of the measured single crystals. (c) Temperature dependence of the heat capacity of LuNb$_6$Sn$_6$. A pronounced anomaly at 84.3 K marks the CDW transition. Inset: zoomed-in view near $T_{\mathrm{CDW}}$ highlighting the transition and the thermal hysteresis between cooling (blue) and heating (red) measurements. The green curve represents heat-capacity data obtained using the 2$\tau$ relaxation method, and the red and blue curves were acquired with the large-pulse single-slope method.}\label{Fig1}
\end{figure}

\begin{figure}
\centering
\includegraphics[width=0.75\linewidth]{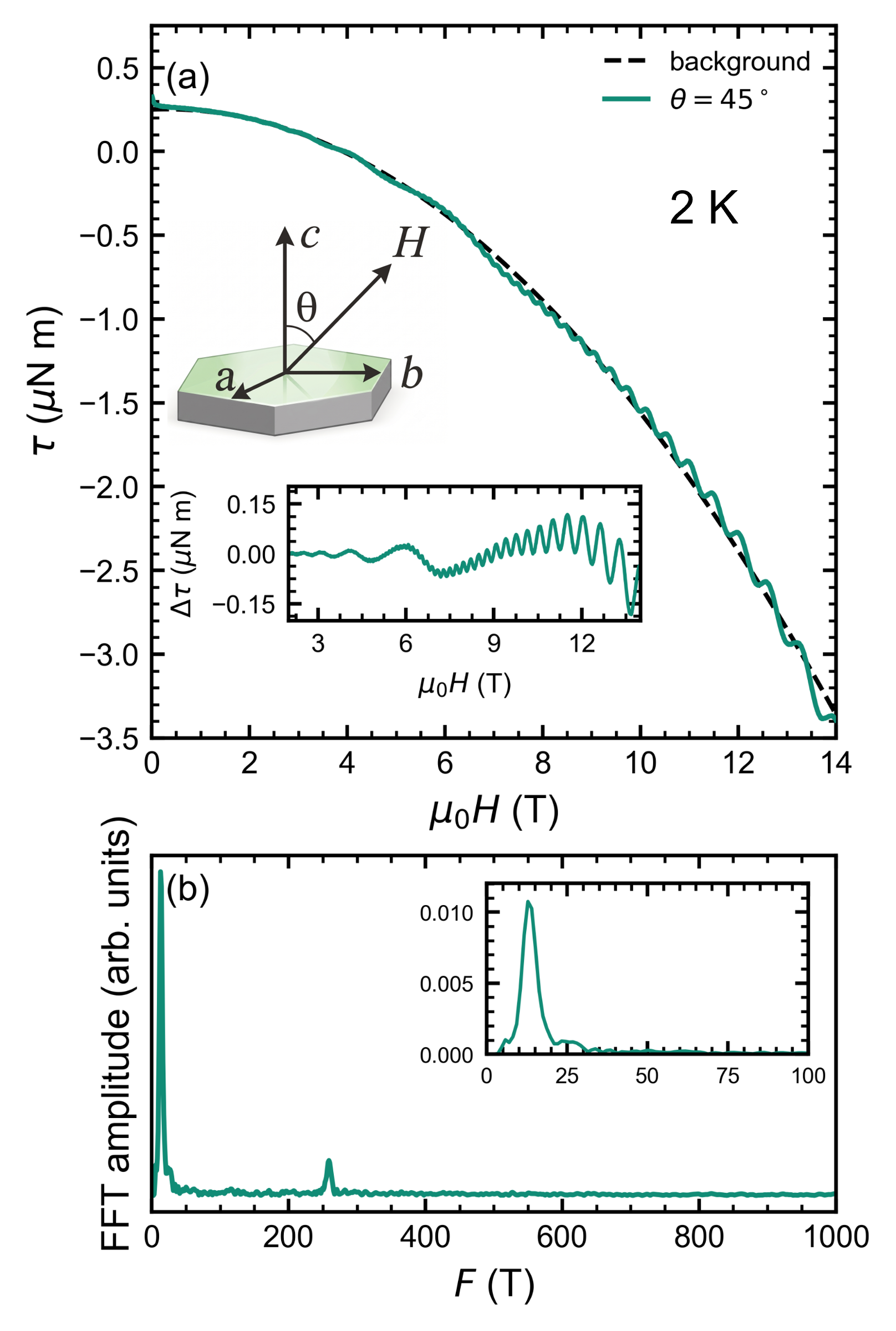}
\caption{(a) Torque $\tau$ as a function of magnetic field $H$ for a LuNb$_6$Sn$_6$ single crystal (Sample S1) measured at 2 K with the field applied at an angle $\theta = 45^\circ$. $\tau(H)$ exhibits clear dHvA oscillations above 2 T. The black, dashed line represents a second-order polynomial fit to the background. Inset: schematic illustration of the tilt angle and background-subtracted torque signal showing the dHvA oscillations. (b) FFT spectrum of the oscillatory signal shown in (a). Two dominant frequencies are observed at $F_\alpha = 12.8$ T and $F_\beta = 259$ T.}\label{Fig2}
\end{figure}

\begin{figure}
\centering
\includegraphics[width=0.8\linewidth]{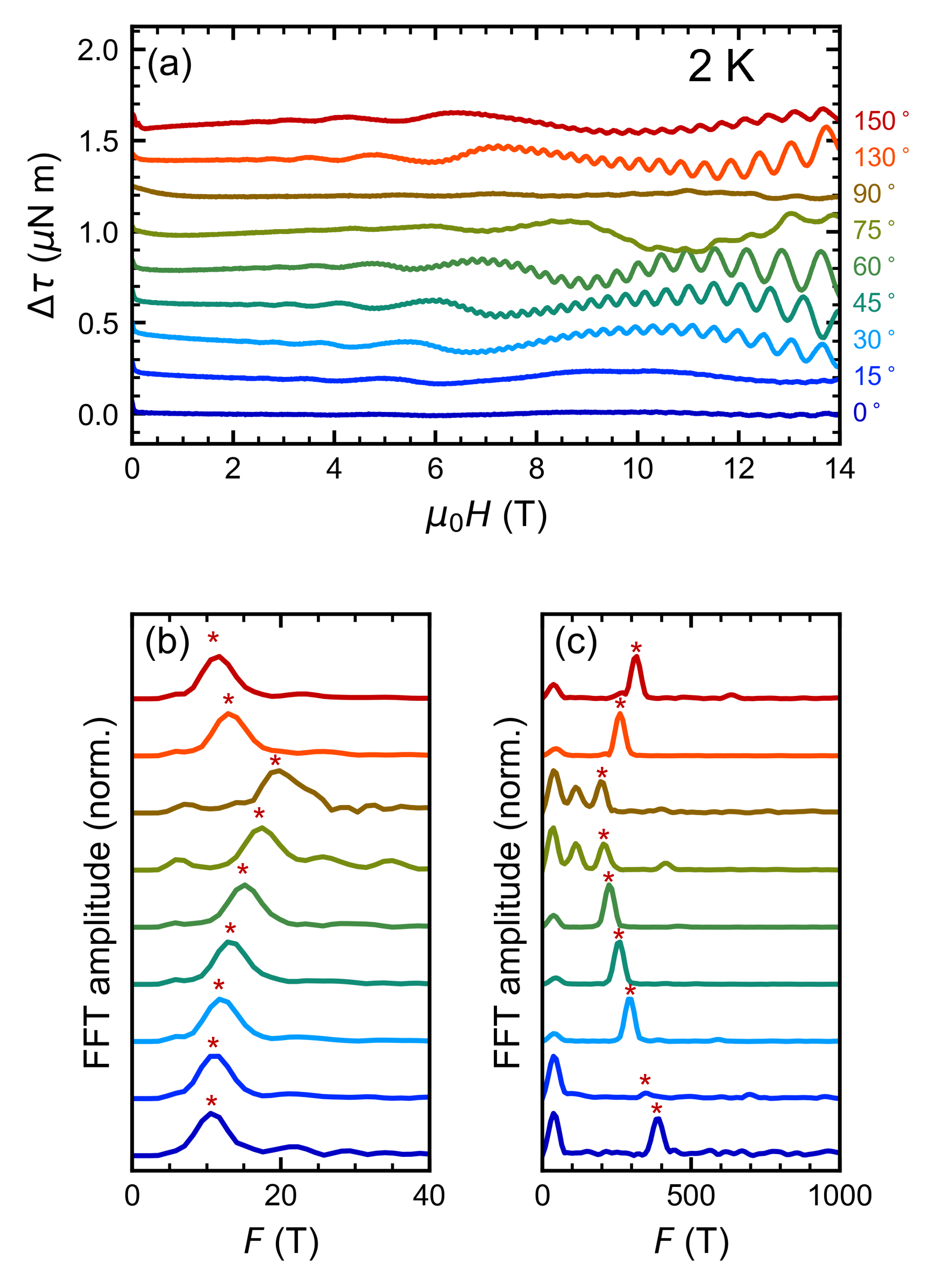}
\caption{(a) Background-subtracted angle-dependent torque data for a LuNb$_6$Sn$_6$ single crystal (Sample S1) at 2 K. dHvA oscillations are observed for all tilt angles. (b) Frequency spectra in the low-frequency range with FFT analysis performed between 2--14 T. (c) Frequency spectra in the high-frequency range with FFT analysis performed between 8--14 T. In both (b) and (c) FFT amplitudes have been normalized by the maximum peak amplitude and offset by a constant to emphasize the angular dependence. Frequencies marked with an asterisk are the focus of the quantitative analysis.}\label{Fig3}
\end{figure}

\begin{figure}
\centering
\includegraphics[width=0.75\linewidth]{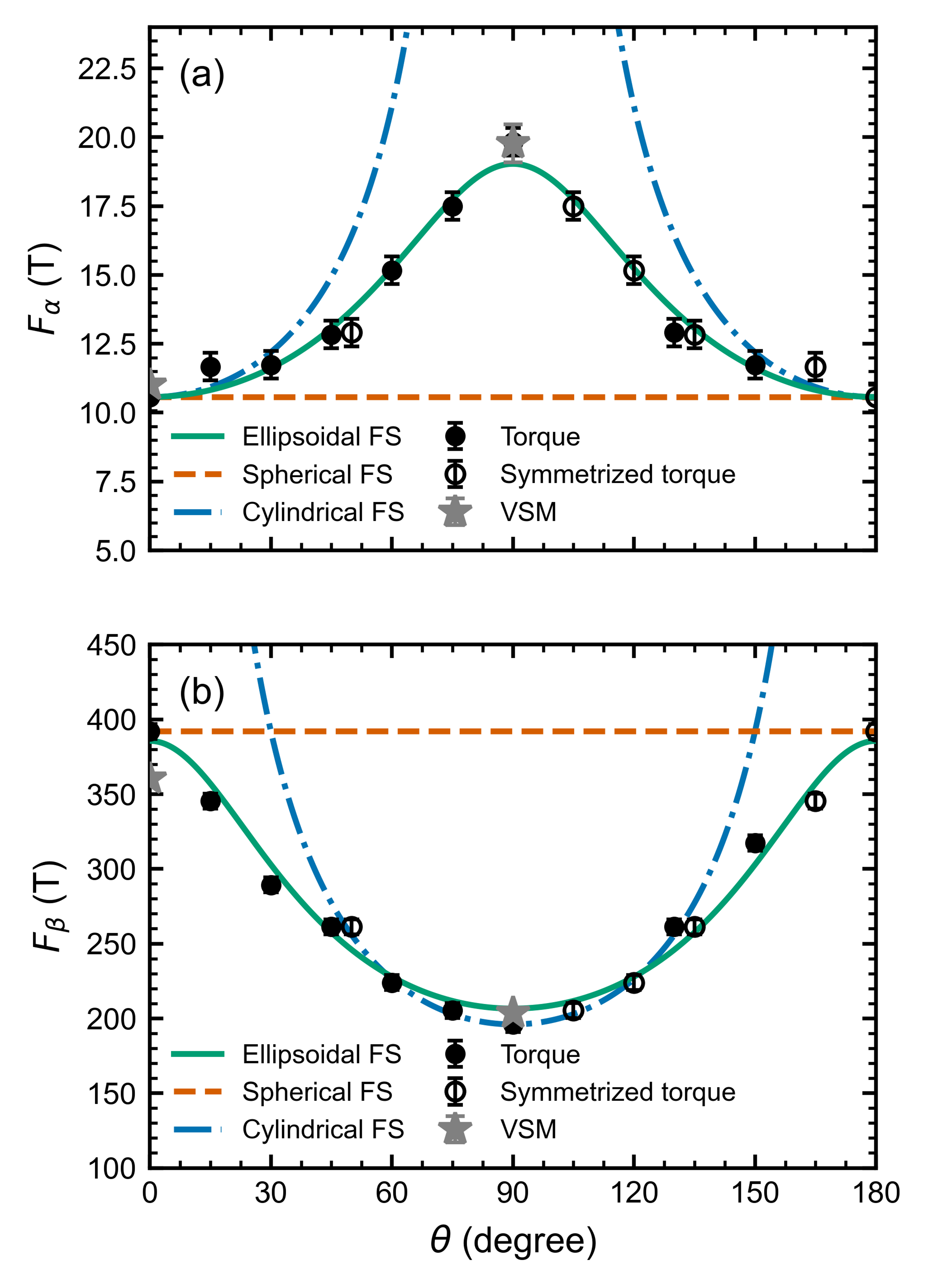}
\caption{Angle dependence of (a) $F_\alpha$ and (b) $F_\beta$ for LuNb$_6$Sn$_6$ (Sample S1). Solid circles represent data obtained from torque measurements and open circles are symmetrized data. Gray stars correspond to VSM measurements. The solid curves in both panels represent the frequency–angle relation expected for an ellipsoidal FS, indicating that both $F_\alpha$ and $F_\beta$ originate from ellipsoidal Fermi pockets. The dashed lines are expected frequency-angle relations for spherical or cylindrical FS. The values and uncertainties of the frequencies in both panels are from fitting a pseudo-Voigt function for each peak. All data acquired at T = 2 K.}\label{Fig4} 
\end{figure}

\begin{figure}
\centering
\includegraphics[width=0.8\linewidth]{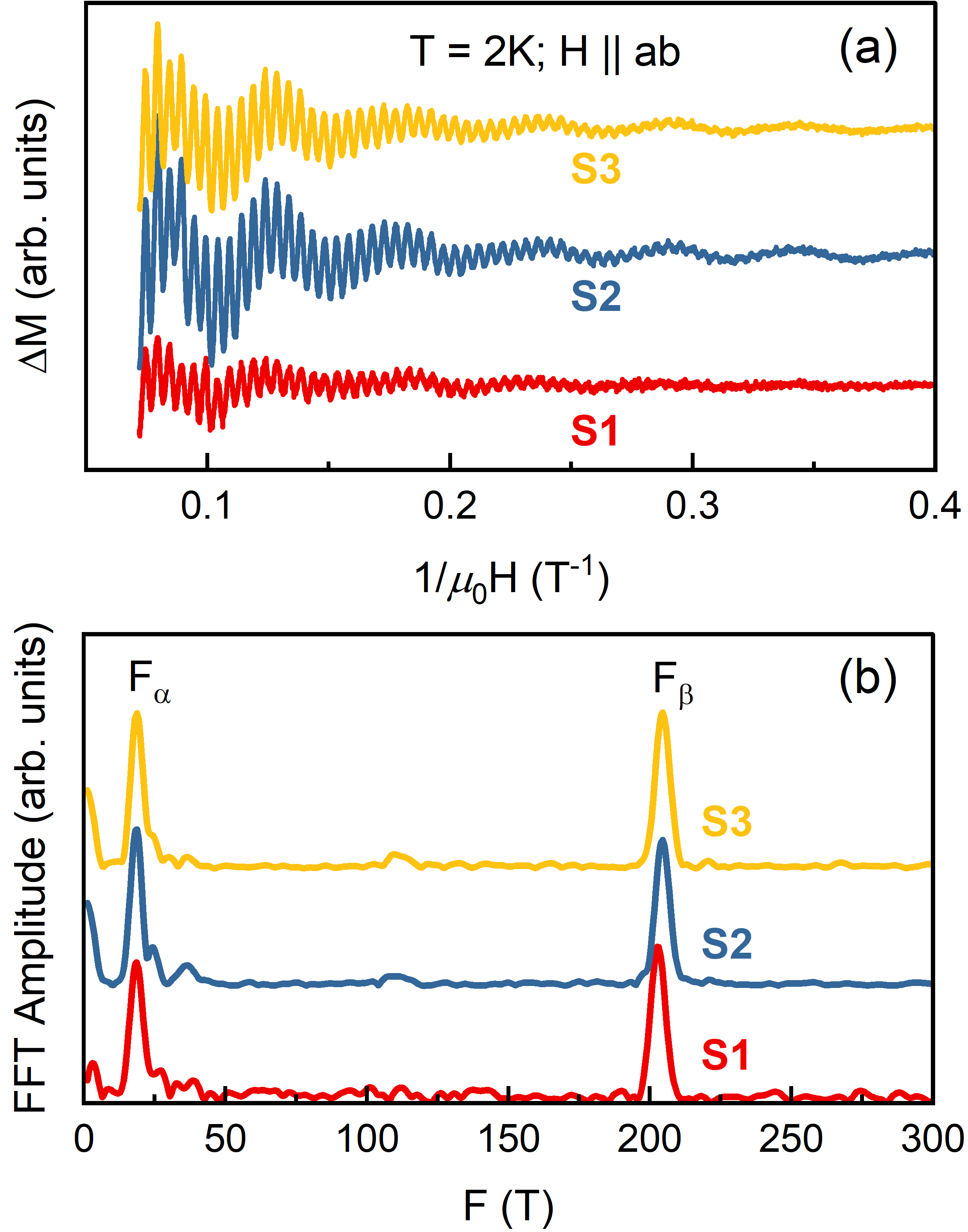}
\caption{(a) Background-subtracted magnetization as a function of inverse magnetic field with $H \parallel ab$ for three LuNb$_6$Sn$_6$ samples (S1, S2, and S3) measured at 2 K. (b) In all three samples, FFTs of the background-subtracted magnetization reveal two main quantum oscillation frequencies near $F_{\alpha} \approx 19\,\mathrm{T}$ and $F_{\beta} \approx 204\,\mathrm{T}$, emphasizing the reproducibility of the results. The magnetization data and FFTs have been offset by constants for visual clarity.}\label{Fig5}  
\end{figure}

\begin{figure}
\centering
\includegraphics[width=0.7\linewidth]{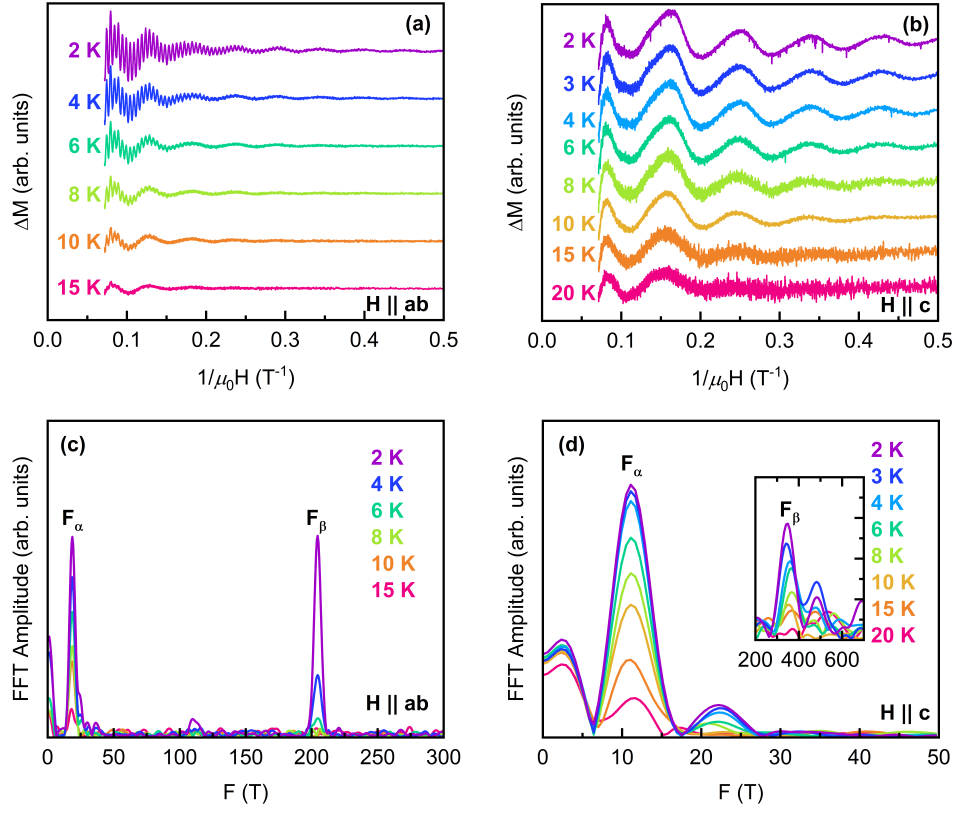}
 \caption{Background-subtracted magnetization as a function of inverse magnetic field for LuNb$_6$Sn$_6$ (Sample S3), measured at different temperatures with the field applied along (a) $H \parallel ab$ and (b) $H \parallel c$. The data have been offset by a constant for visual clarity. Clear dHvA oscillations are observed in both field orientations. (c,d) FFT of the oscillatory signals for $H \parallel ab$ and $H \parallel c$, respectively. For $H \parallel ab$, two main quantum oscillation frequencies are observed at $F_{\alpha} = 19\,\mathrm{T}$ and $F_{\beta} = 204\,\mathrm{T}$. When the magnetic field is applied to $H \parallel c$, these frequencies shift to $11\,\mathrm{T}$ and $361\,\mathrm{T}$, respectively. Inset in (d): Zoomed-in view of the FFT highlighting the $F_{\beta}$ peak for $H \parallel c$.}\label{Fig6}
\end{figure}

\begin{figure}
\centering
\includegraphics[width=0.8\linewidth]{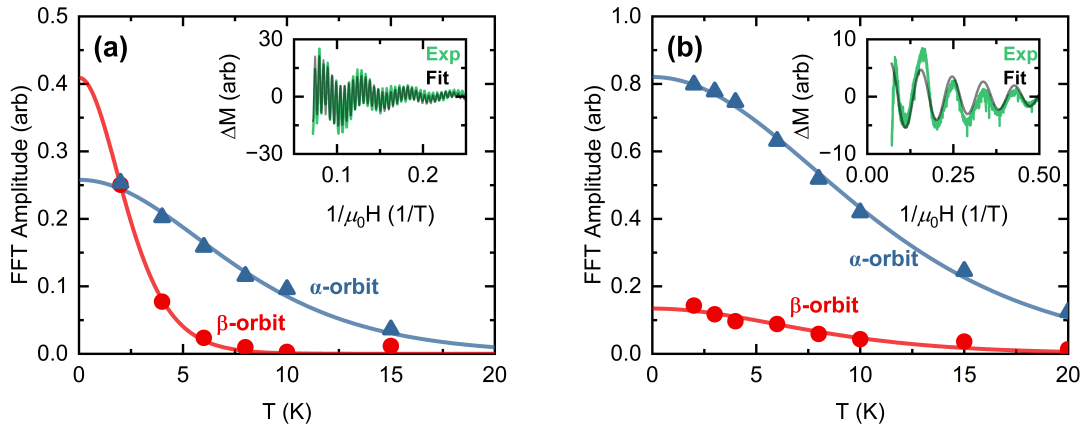}
\caption{LK analysis of LuNb$_6$Sn$_6$ (Sample S3) based on dHvA oscillations observed in magnetization. Temperature dependence of the FFT amplitudes corresponding to the $F_{\alpha}$ and $F_{\beta}$ frequencies for (a) $H \parallel ab$ and (b) $H \parallel c$. The solid curves represent the best fits to the data using the LK formula (Eq.~\ref{LK}). Insets in both panels show LK fitting as a function of magnetic field to determine $T_D$ at 2 K.}\label{Fig7}
\end{figure}

\begin{figure}
\centering
\includegraphics[width=0.8\linewidth]{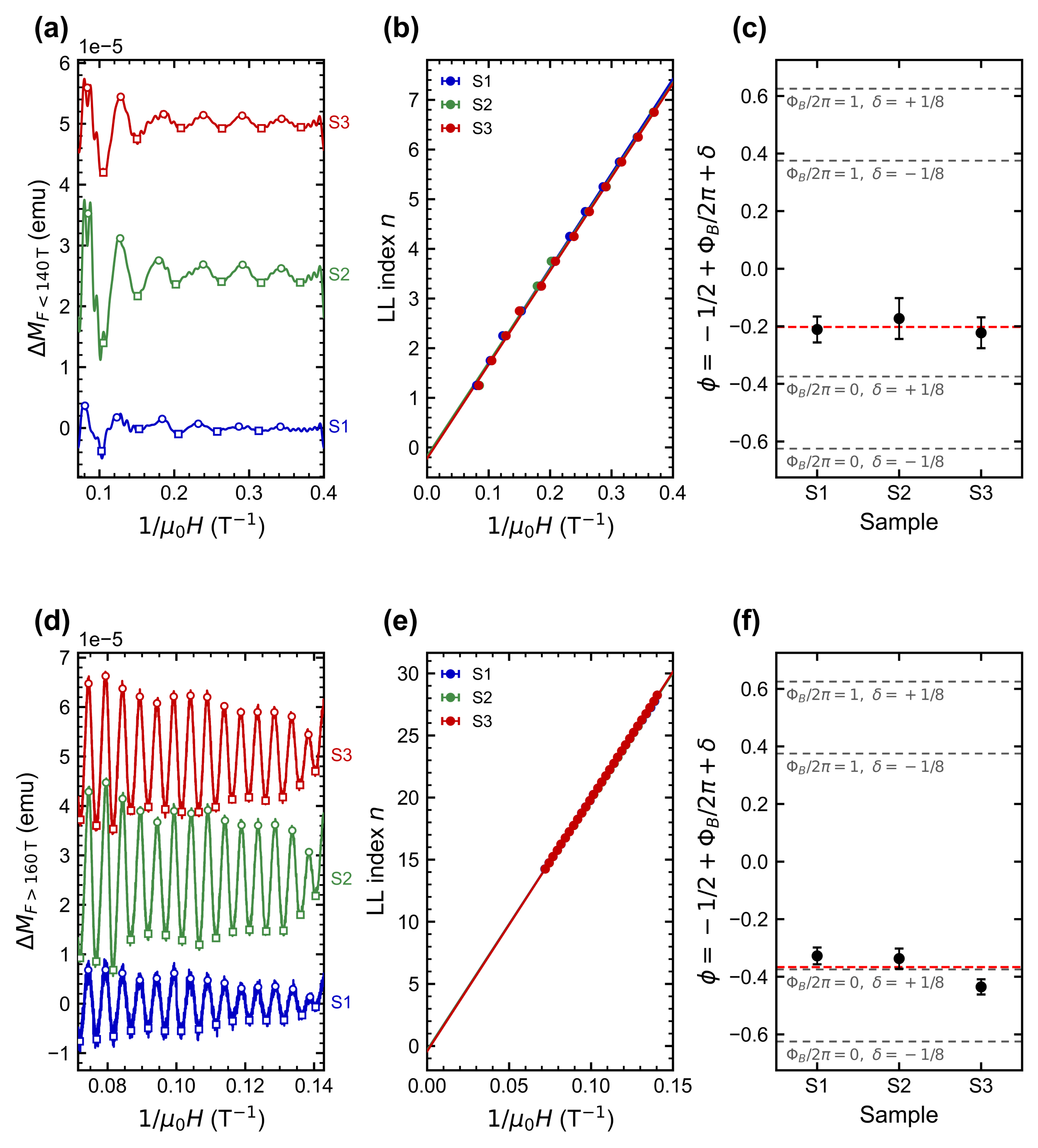}
\caption{Landau level fan diagram analysis for three LuNb$_6$Sn$_6$ samples for H $\parallel$ ab. (a,d) Oscillatory magnetization as a function of inverse magnetic field for low-frequency ($F<140$ T) and high-frequency ($F>160$ T) oscillations, respectively. Open symbols mark the extrema used for Landau-level indexing. (b,e) Corresponding Landau fan diagrams. Symbols are the experimental oscillation extrema positions, and solid lines are linear fits used to extract the phase. (c,f) Extracted phase, $\phi=-1/2+\Phi_B/2\pi+\delta$, for each sample. The horizontal gray lines indicate the expected values for trivial ($\Phi_B=0$ and $\Phi_B=2\pi$) Berry phases for $\delta=\pm1/8$, while the red dashed line denotes the average phase over the three samples. Error bars represent the uncertainties from the linear fits.}\label{Fig8}
\end{figure}

\begin{figure}
\centering
\includegraphics[width=1.0\linewidth]{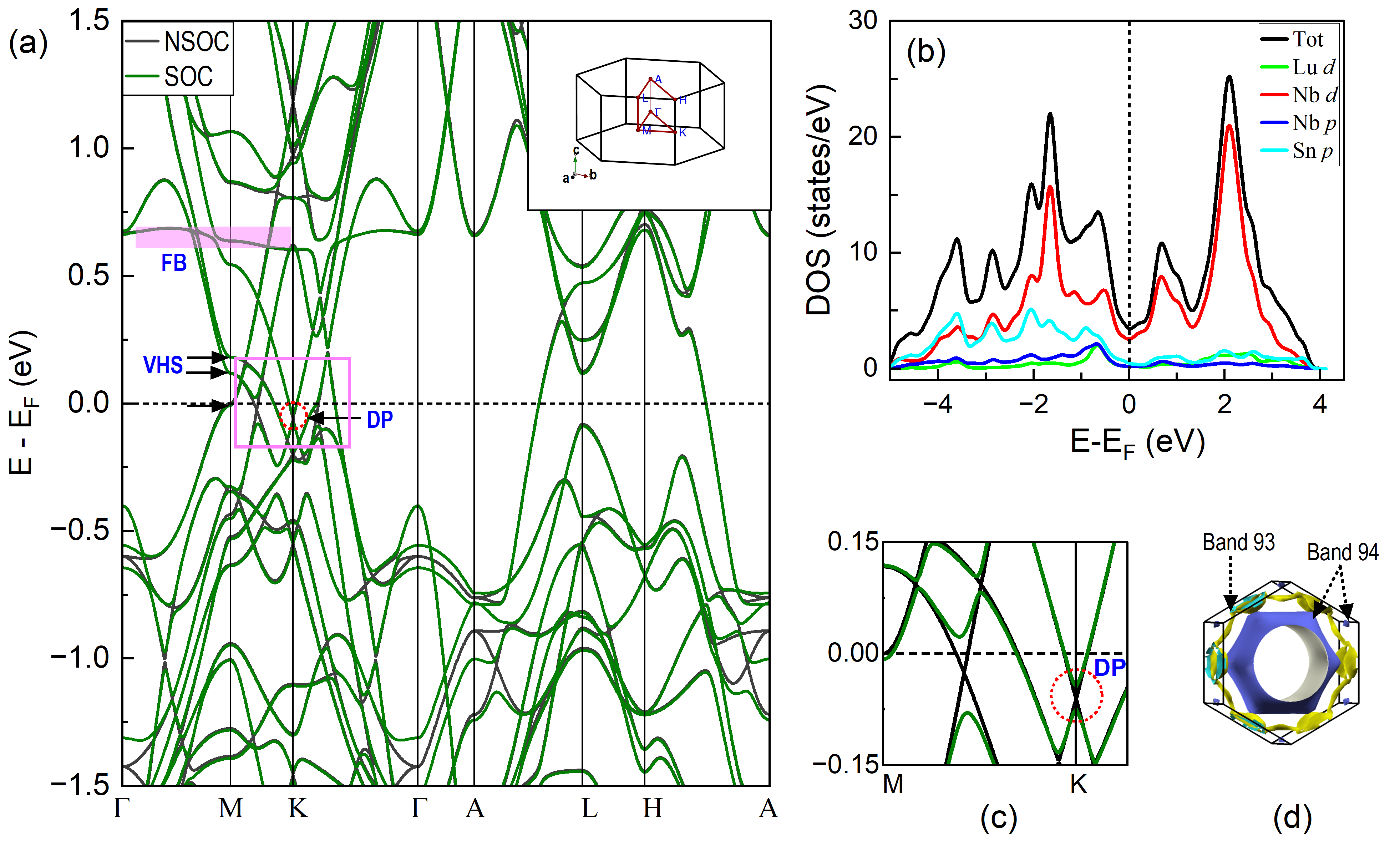}
\caption{(a) Electronic band structure of LuNb$_6$Sn$_6$ calculated with and without SOC using DFT. Several notable features  near the $E_F$ are identified, including a FB, VHS, and DP, highlighted by the shaded region, arrows, and dotted circle, respectively. Inset: Hexagonal Brillouin zone of LuNb$_6$Sn$_6$, showing the high-symmetry points used in the band structure calculations. (b) Orbital-resolved DOS of LuNb$_6$Sn$_6$. The electronic states near the $E_F$ are predominantly contributed by Nb-$d$ orbitals. (c) Enlarged band structure of the boxed region in (a), showing that SOC opens a $\sim$120 meV gap at the DP. (d) Band-resolved FS of LuNb$_6$Sn$_6$ using the DFT-computed $E_F$, consisting of a warped ellipsoidal Fermi pocket centered at the $\Gamma$ point (Band 94) and a chain-like FS sheet at the Brillouin zone boundary (Band 93).}\label{Fig9}
\end{figure}

\begin{figure}
\centering
\includegraphics[width=1.0\linewidth]{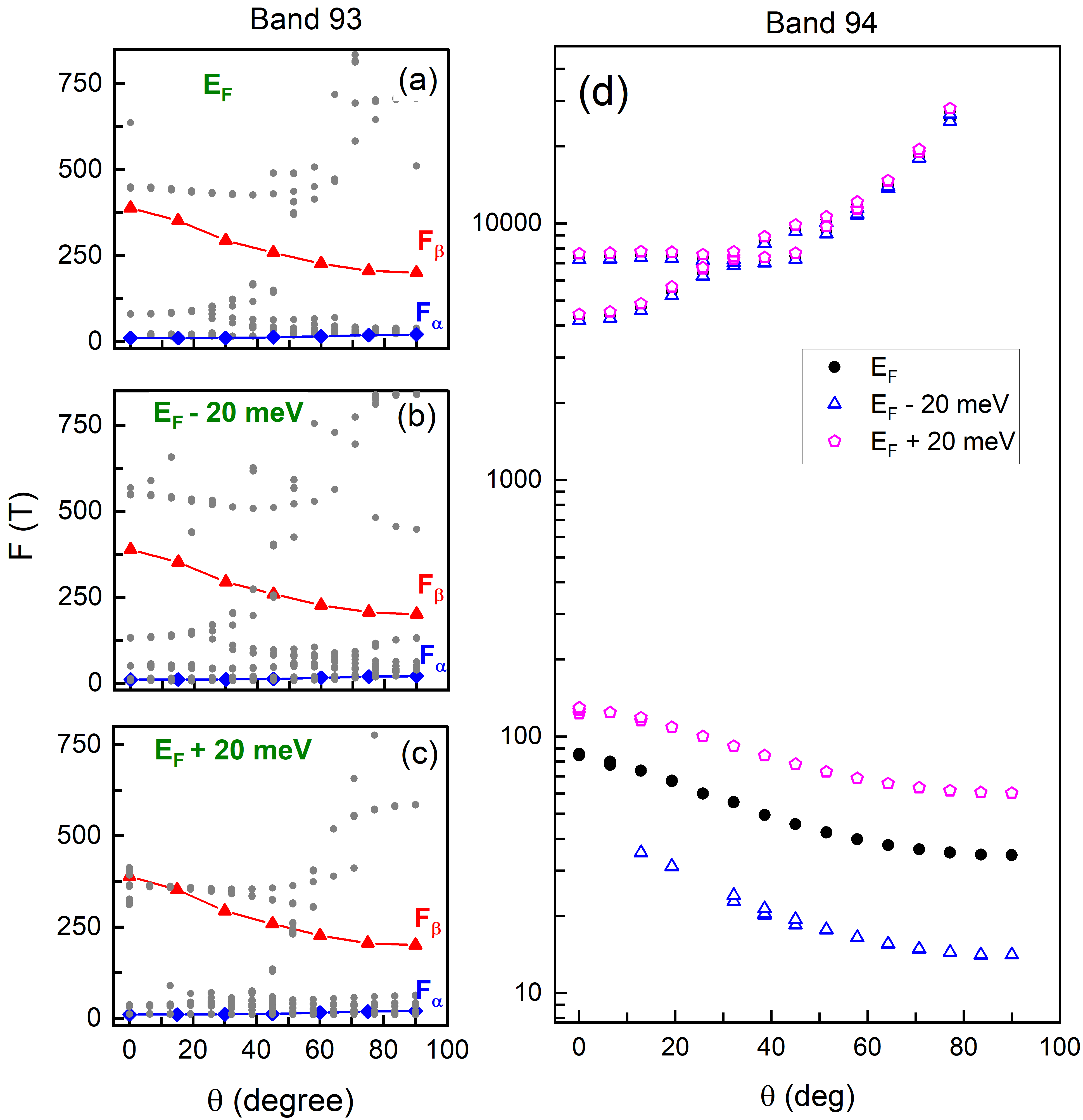}
\caption{Calculated quantum oscillation frequencies from the FS pocket of Band~93 with (a) no Fermi-level shift, (b) $E_F-20$~meV, and (c) $E_F+20$~meV. The magnetic field was rotated from the c-axis toward the a-axis direction within the ac-plane. The experimentally observed dHvA frequencies ($F_\alpha$ and $F_\beta$) are included for comparison. While the calculated frequencies with $E_F+20$~meV are in good agreement with $F_\beta$ at $\theta = 0^\circ$, they deviate from the experimental angular dependence at higher angles. (d) Calculated quantum oscillation frequencies from the FS pocket of Band~94 for different Fermi-level shifts. The Fermi-level shift modifies the calculated frequencies for both bands. The y-axis in (d) is plotted on a logarithmic scale for clarity.}\label{Fig10} 
\end{figure}

\end{document}